\documentclass[paper]{JHEP3}
\pdfoutput=1
\usepackage{amsmath,amssymb,amsthm,amscd,graphicx}
\input epsf.sty

\theoremstyle{definition}

\newcommand{\CC}{{\cal C}}

\newcommand{\CI}{{\cal I}}

\newcommand{\CM}{{\cal M}}
\newcommand{\CN}{{\cal N}}
\newcommand{\CO}{{\cal O}}

\newcommand{\CR}{{\cal R}}

\newcommand{\CW}{{\cal W}}

\def\IZ{{\mathbb Z}}
\def\IR{{\mathbb R}}
\def\IC{{\mathbb C}}
\def\IP{{\mathbb P}}

\def\IS{{\mathbb S}}

\def\IH{{\mathbb H}}

\newcommand{\tr}{{\rm Tr}}
\newcommand{\re}{{\rm e}}
\newcommand{\ri}{{\rm i}}
\newcommand{\rd}{{\rm d}}
\renewcommand{\d}{\partial}

\newcommand{\mon}{\mathop{\rm mon}}
\newcommand{\sign}{\mathop{\rm sign}}
\renewcommand{\d}{\partial}

\newcommand{\trop}{{\mathrm{trop}}}
\newcommand{\pure}{{\mathrm{p}}}
\newcommand{\matter}{{\mathrm{m}}}

\newcommand{\be}{\begin{equation}}
\newcommand{\ee}{\end{equation}}
\newcommand{\ba}{\begin{aligned}}
\newcommand{\ea}{\end{aligned}}
\newcommand{\ben}{\begin{eqnarray}\displaystyle}
\newcommand{\een}{\end{eqnarray}}

\newcommand{\sectiono}[1]{\section{#1}\setcounter{equation}{0}}

\newdimen\tableauside\tableauside=1.0ex
\newdimen\tableaurule\tableaurule=0.4pt
\newdimen\tableaustep
\def\phantomhrule#1{\hbox{\vbox to0pt{\hrule height\tableaurule width#1\vss}}}
\def\phantomvrule#1{\vbox{\hbox to0pt{\vrule width\tableaurule height#1\hss}}}
\def\sqr{\vbox{%
  \phantomhrule\tableaustep
  \hbox{\phantomvrule\tableaustep\kern\tableaustep\phantomvrule\tableaustep}%
  \hbox{\vbox{\phantomhrule\tableauside}\kern-\tableaurule}}}
\def\squares#1{\hbox{\count0=#1\noindent\loop\sqr
  \advance\count0 by-1 \ifnum\count0>0\repeat}}
\def\tableau#1{\vcenter{\offinterlineskip
  \tableaustep=\tableauside\advance\tableaustep by-\tableaurule
  \kern\normallineskip\hbox
    {\kern\normallineskip\vbox
      {\gettableau#1 0 }%
     \kern\normallineskip\kern\tableaurule}%
  \kern\normallineskip\kern\tableaurule}}
\def\gettableau#1{\ifnum#1=0\let\next=\null\else
\squares{#1}\let\next=\gettableau\fi\next}

\newcommand{\figref}[1]{Fig.~\protect\ref{#1}}

\title{Unquenched flavor and tropical geometry in strongly coupled Chern--Simons--matter theories}

\author{
Ricardo Couso Santamar\'\i a$^{a}$, Marcos Mari\~no$^{b,c}$ and Pavel Putrov$^c$
\\
$^a$Departamento de F\'\i sica de Part\'\i culas and IGFAE, 
Universidade de Santiago de Compostela, \\
E-15782, Santiago de Compostela, Spain\\
\\
$^b$D\'epartement de Physique Th\'eorique et $^c$Section de Math\'ematiques,\\
Universit\'e de Gen\`eve, Gen\`eve, CH-1211 Switzerland\\
\\
\email{ricardo.couso@usc.es}, \quad 
\email{marcos.marino@unige.ch}, \quad
\email{pavel.putrov@unige.ch}
}

\abstract{We study various aspects of the matrix models calculating free energies and Wilson loop observables in supersymmetric Chern--Simons--matter theories on the three-sphere. We first develop techniques to extract strong coupling results directly from the spectral curve 
describing the large $N$ master field. We show that the strong coupling limit 
of the gauge theory corresponds to the so-called tropical limit of the spectral curve. In this limit, the curve degenerates to a planar graph, and matrix model calculations reduce to elementary line integrals along the graph. As an important physical application of these tropical techniques, we study $\CN=3$ theories with fundamental matter, both in the quenched and in the unquenched regimes. We calculate the 
exact spectral curve in the Veneziano limit, and we evaluate the planar free energy and Wilson loop observables at strong coupling by using tropical 
geometry. The results are in agreement with the predictions of the AdS duals involving tri-Sasakian manifolds.}    

\begin{document}

\sectiono{Introduction}
Recently, the localization methods of \cite{pestun} were extended to supersymmetric Chern--Simons--matter theories in \cite{kapustin,ktwo,dt}. 
These methods provide matrix model 
representations for the Euclidean partition function of these theories on the three-sphere, as well as of BPS Wilson loop VEVs. In the case of 
the ABJM theory constructed in \cite{abjm, abj}, the resulting matrix models were solved at all orders in the $1/N$ expansion \cite{mp,dmp}
by using previous results \cite{akmv,hy,hkr} on Chern--Simons matrix models \cite{mm} and their large $N$ topological string duals. 
This exact solution made possible the calculation of exact interpolating functions in ABJM theory, testing in this way various strong coupling 
predictions based on its AdS$_4$ dual. In particular, the $N^{3/2}$ behavior of the number of degrees of freedom of the M2 brane theory \cite{kt} 
was derived in \cite{dmp} by calculating the exact planar free energy on the three-sphere. 

The Chern--Simons--matter quiver theory of ABJM has been generalized in many directions, by considering for example more complicated 
quivers or by adding flavor, i.e. matter multiplets in the fundamental representation. The resulting theories have conjectural AdS duals which lead to predictions for the strong coupling behavior of various observables. If the Chern--Simons--matter theories have enough supersymmetry, the methods of \cite{kapustin} provide matrix model representations for the partition function on the three-sphere and for Wilson loop VEVs.
In general, these matrix models are quite complicated, and solving them for all values of the coupling, even in the planar limit, 
is a difficult task. In this respect, the ABJM theory is rather exceptional since it can be related to a well-known topological string theory, and this 
is what made possible to obtain very explicit interpolating functions \cite{mp,dmp}. In some more general cases, like the one considered in \cite{suyama}, one can 
write down the planar resolvent in terms of contour integrals, but the resulting expressions are not explicit enough to extract in a straightforward way the physical information on the observables. 

As a matter of fact, finding an exact solution to the matrix model at all values of the coupling might be a Pyrrhic victory: 
if we just want to test the AdS predictions for a given Chern--Simons--matter theory, we simply need the leading result at strong coupling. 
Therefore, we would like to have a method to perform 
calculations in the matrix model directly in the strong coupling limit, in the same way that we 
can use matrix model perturbation theory to obtain results in the weak coupling limit.  

The first goal of this paper is to develop techniques to perform such strong coupling computations. The basic idea is that, as emphasized in \cite{dmp}, the strong coupling behavior of the ABJM theory corresponds to the limit of large complex structure for the spectral curve describing the 
large $N$ distribution of eigenvalues. It turns out that this limit has a number of simplifying features that have been much studied recently in the mathematical literature: it is the so-called {\it tropical limit} of the curve. In this limit, the spectral curve degenerates to a two-dimensional graph, and the calculation of periods reduces to the computation of line integrals along this graph. Many recent results in tropical geometry can in fact be adapted to our purposes, and we obtain in this way simple methods to evaluate the strong coupling limit of observables in ABJM theory. 

The main reason to develop techniques based on tropical geometry is of course to be able to go beyond ABJM theory and obtain strong coupling 
results for more complicated Chern--Simons--matter theories from their matrix models. In this paper we will focus on the theories with fundamental matter introduced in \cite{hk,gj,taka}. These theories have only $\CN=3$ supersymmetry, and they are simple toy models for the incorporation of flavors in the 
AdS/CFT correspondence. As in other theories with fundamental matter, one can study them in the quenched approximation, or one can consider the full unquenched regime, in which matter loops are not suppressed. We show that both approaches can be incorporated in the matrix models of \cite{kapustin}. The quenched approximation leads to a systematic expansion in powers of the number of flavors, where each term can be 
computed as a correlator in ABJM theory. In order to study the unquenched theory we find the exact planar resolvent of the $\CN=3$ theory in the 
Veneziano limit \cite{veneziano}. In both cases, quenched and unquenched, we can use tropical techniques to calculate 
observables at strong coupling, testing in this way the 
AdS predictions for these theories made in \cite{hk,gj,taka}. In particular, we reconstruct from the matrix model calculation the full volume of the 
tri-Sasaki Einstein manifold appearing in their M-theory dual. 

The organization of this paper is as follows. In section 2 we review the relevant Chern--Simons--matter theories, their matrix model description, as well as their AdS duals. In section 3 we review the exact solution of the ABJM theory matrix model worked out in \cite{mp,dmp}, and we show how 
to perform direct strong coupling calculations by using the tropical limit of the exact solution. In section 4 we start our study of the $\CN=3$ Chern--Simons--matter theories with flavors and we present the matrix model analogue of the quenched expansion. We also evaluate the first terms in this 
expansion at strong coupling by developing techniques based on tropical geometry. In section 5 we solve the $\CN=3$ theory in the Veneziano limit, we test the exact planar resolvent at weak coupling, and we calculate the free energy and Wilson loop VEVs at strong coupling, recovering in this way the 
AdS predictions. Finally, in section 6 we state our conclusions and open problems. 

When the first version of this paper was being typed, the very interesting paper \cite{hklebanov} appeared, which has some overlap with our results. 
One of the goals of \cite{hklebanov} is also to obtain results directly at strong coupling in the matrix model, and they develop a simple and powerful technique 
to do that based on a careful analysis of the eigenvalue density. Their main conclusion, that the 
density of eigenvalues is piecewise linear in this regime, corresponds precisely to the tropical limit of the spectral curves considered in this paper. In the Appendix we 
use the methods of \cite{hklebanov} to re-derive the free energy of the $\CN=3$ theory 
with fundamental matter, and we clarify the relationship between their approach and the 
tropical techniques of this paper. 

 \sectiono{Supersymmetric Chern--Simons--matter theories and matrix models}

\subsection{ABJM theory and matrix models}

The ABJM theory \cite{abjm,abj} is a supersymmetric 
quiver Chern--Simons--matter theory in three dimensions with gauge group $U(N_1)_k \times U(N_2)_{-k}$ and $\CN=6$ supersymmetry. 
The Chern--Simons actions have couplings $k$ and $-k$, respectively. The theory contains in addition four hypermultiplets: there are two bifundamental chiral fields 
$(A_1, A_2)$ in the representation $(N_1, \overline N_2)$, and two bifundamental chiral fields 
$(B_1, B_2)$ in the representation $(\overline N_1, N_2)$. 

The ABJM matrix model, obtained in \cite{kapustin}, gives an explicit integral expression for the partition function of the ABJM theory on $\IS^3$, as well as for 
Wilson loop VEVs. This matrix model is defined by the partition function 
\be
\label{kapmm}
\ba
&Z_\text{ABJM}(N_1, N_2, g_s)\\
&={1\over N_1! N_2!} \int \prod_{i=1}^{N_1}{ \rd \mu_i  \over 2\pi} \prod_{j=1}^{N_2} {\rd \nu_j \over 2\pi}
 {\prod_{i<j} \left( 2 \sinh \left( {\mu_i -\mu_j \over 2}\right) \right)^2 \left(2 \sinh \left( {\nu_i -\nu_j \over 2}\right) \right)^2 \over 
\prod_{i,j}  \left(2 \cosh \left( {\mu_i -\nu_j \over 2}\right) \right)^2} \re^{-{1\over 2g_s}\left(  \sum_i \mu_i^2 -\sum_j \nu_j^2\right)}, 
\ea
\ee
where the coupling $g_s$ is related to the Chern--Simons coupling $k$ of the ABJM theory as
\be
g_s={2 \pi \ri \over k}.
\label{gs}
\ee
The normalization of this matrix integral has been explained in \cite{dmp}. It was shown in \cite{mp} that 
the ABJM matrix model is closely related to the $L(2,1)$ lens space matrix model introduced in \cite{mm} and further studied in \cite{akmv,hy,hoy}. 
This matrix model is defined by the partition function
\be
\label{intdef}
\ba
Z_{L(2,1)}(N_1, N_2, g_s)={1\over N_1! N_2!} \int \prod_{i=1}^{N_1}{ \rd \mu_i  \over 2\pi} \prod_{j=1}^{N_2} {\rd \nu_j \over 2\pi} & \prod_{i<j} \left( 2 \sinh  \left( {\mu_i -\mu_j \over 2}\right) \right)^2 \left( 2 \sinh  \left( {\nu_i -\nu_j \over 2}\right) \right)^2\\
\times &\prod_{i,j} \left( 2  \cosh  \left( {\mu_i -\nu_j \over 2}\right) \right)^2\,  \re^{-{1\over 2g_s}\left(  \sum_i \mu_i^2 +\sum_j \nu_j^2\right)}.
\ea
\ee
The relation between the partition functions is simply
\be
\label{changesignZ}
Z_\text{ABJM}(N_1, N_2, g_s) =Z_{L(2,1)}(N_1, -N_2, g_s). 
\ee
Since the large $N$ expansion of the free energy gives a sequence of analytic functions of $N_1$, $N_2$, once these functions are known in one model, they can be obtained in the other 
by the trivial change of sign $N_2 \rightarrow -N_2$. Equivalently, it can be shown \cite{mp,dt} that the ABJM matrix model is the supergroup version of the lens space matrix model of \cite{mm,akmv}.

Natural observables in this theory are supersymmetric Wilson loop operators. The $1/6$ BPS Wilson loop has been constructed in \cite{dp,cw,rey} and 
it is labelled by a representation $R$ of $U(N_1)$. One of the main results of \cite{kapustin} is that the VEV of this Wilson loop in ABJM theory
can be obtained by calculating the normalized VEV of the matrix $\re^{\mu_i}$ 
in the matrix model (\ref{kapmm}), {\em i.e.}, 
\be
\label{16WL}
\langle W^{1/6}_R\rangle
=g_s \left\langle \tr_R\left (\re^{\mu_i}\right)  \right\rangle_\text{ABJM},
\ee
A 1/2 BPS loop $W^{1/2}_\CR$ was constructed in \cite{dt} , where $\CR$ is a 
representation of the supergroup $U(N_1|N_2)$. In \cite{dt} it was also shown 
that it localizes to the matrix model correlator in the ABJM matrix model
\be
\label{12wl}
\langle W^{1/2}_\CR \rangle
=g_s \left\langle \text{Str}_\CR\, U \right\rangle_\text{ABJM},
\ee
where $U$ is the diagonal matrix
\be
\label{U}
U=\begin{pmatrix}\re^{\mu_i} &0 \\ 0& -\re^{\nu_j} \end{pmatrix}.
\ee

\subsection{Flavored theory and its gravity dual}

It is possible to flavor the ABJM theory by adding matter hypermultiplets in the fundamental representation \cite{hk,gj,taka}. More precisely, one adds $N_f^{(i)}$ multiplets $(Q_i, \widetilde Q_i)$, with $i=1,2$. The fields $Q_i$, $i=1,2$ are in the representations $(N_1, 1)$ and $(1, N_2)$, respectively, while $\widetilde Q_i$ are in the conjugate representations 
$(\overline N_1, 1) $ and $(1, \overline N_2)$, respectively. This matter content breaks the $\CN=6$ supersymmetry of the ABJM theory down to $\CN=3$. Notice that the ABJM theory can be obtained, formally, as the limit 
\be
N_f^{(i)}\rightarrow 0
\ee
of the flavored theory. We will denote by
\be
N_f =N_f^{(1)} +N_f^{(2)}
\ee
the total number of flavours.

The partition function of this $\CN=3$ theory can be also computed in terms of matrix models, as explained in \cite{kapustin}. The inclusion of extra matter hypermultiplets just 
leads to the insertion of determinant-type operators in the matrix integral (\ref{kapmm}):
\be
\label{gjmm}
\ba
&Z_{\CN=3}(N_1, N_2, N_f ^{(1)}, N_f^{(2)}, g_s)\\
&={1\over N_1! N_2!} \int \prod_{i=1}^{N_1}{ \rd \mu_i  \over 2\pi} \prod_{j=1}^{N_2} {\rd \nu_j \over 2\pi}
 {\prod_{i<j} \left( 2 \sinh \left( {\mu_i -\mu_j \over 2}\right) \right)^2 \left(2 \sinh \left( {\nu_i -\nu_j \over 2}\right) \right)^2 \over 
\prod_{i,j}  \left(2 \cosh \left( {\mu_i -\nu_j \over 2}\right) \right)^2}  \\
& \qquad  \times \prod_{i=1}^{N_1} \left(  2 \cosh {\mu_i \over 2} \right)^{-N_f^{(1)}} \prod_{i=1}^{N_2} \left(  2 \cosh {\nu_i \over 2} \right)^{-N_f^{(2)}} \re^{-{1\over 2g_s}\left(  \sum_i \mu_i^2 -\sum_j \nu_j^2\right)}.
\ea
\ee

The large $N$ dual of the ABJM Chern--Simons--matter theory is given by type IIA string theory on AdS$_4 \times \IC \IP^3$, which lifts to M-theory on AdS$_4 \times \IS^7/\IZ_k$ \cite{abjm,abj}. It was conjectured in \cite{hk,gj,taka} that, when $N_1=N_2=N$, the $\CN=3$ theory with flavor has a type IIA large $N$ dual where $N_f$ D6 branes wrap the $\IR \IP^3$ cycle inside $\IC \IP^3$. This is the four-dimensional counterpart of the original construction of \cite{kk}, which adds flavor to AdS$_5$ by wrapping D7 branes around an $\IS^3$ inside $\IS^5$. The flavored $\CN=3$ theory also describes $N$ M2 branes probing an eight-dimensional hyperK\"ahler cone $\CM_8$ with ${\rm Sp}(2)$ holonomy. The base of this cone is a 
tri-Sasakian space $X_7$. The space $\CM_8$ is a particular member of a family of hyperK\"ahler cones 
$\CM_8({\bf t})$ labeled by three natural numbers ${\bf t}=(t_1, t_2, t_3)$. These cones can be constructed as hyperK\"ahler quotients 
\be
\IH^3 /// U(1),
\ee
where the $U(1)$ action is characterized by the three charges ${\bf t}$. The bases $X_7({\bf t})$ of these cones give an infinite family of tri-Sasakian manifolds known as Eschenburg spaces, see \cite{ly} for a detailed study and references to the relevant literature. The dual to the $\CN=3$ Chern--Simons--matter theory with a total number of $N_f$ fundamentals has charges
\be
\label{partt}
{\bf t}=\left( N_f, N_f, k\right). 
\ee
In the following, the eight-dimensional cone corresponding to this charge will be simply denoted by $\CM_8$. 

At large $N$ the above theory of $N$ M2 branes is described by M-theory on the manifold
\be
\text{AdS}_4 \times X_7, 
\ee
where $X_7$ is the tri-Sasakian seven manifold corresponding to (\ref{partt}). This background is the eleven-dimensional lift of the type IIA background with $N_f$ D6 branes. The eleven-dimensional metric and four-form flux are given by
\be
\ba
\rd s_{11}^2 &= R_{X_7}^2\left( \frac{1}{4}\rd s_{\text{AdS}_4}^2+\rd s_{X_7}^2\right), \\
	G &= \frac{3}{8}R_{X_7}^3 \omega_{\text{AdS}_4},
\ea
\ee
where $ \omega_{\text{AdS}_4}$ is the volume form with unit radius. 
The radius $R_{X_7}$ is determined by the flux quantization condition
\be
(2\pi \ell_p)^6 N=\int_{C_7} \star_{11} G =6 R_{X_7}^6 {\rm vol}(X_7).
\ee
In this equation, $\ell_p$ is the eleven-dimensional Planck length, $C_7$ is a	cycle enclosing the brane and homologous 
to $X_7$, and the volume of $X_7$ (with unit radius) is given by \cite{ly}
\be
\label{xitri}
{\rm vol}(X_7)={{\rm vol}(\IS^7) \over k \, \xi^2 \left(\mu\right)},
\ee
where
\be
\label{muxidef}
\xi(\mu)={1+\mu \over {\sqrt{1+\mu/2}}}, \qquad \mu={N_f \over k}. 
\ee
Newton's constant in four dimensions can be obtained by standard compactification of the Einstein--Hilbert action in eleven dimensions, 
\be
\label{newnew}
\frac{1}{16\pi G_{4}}=\frac{k^{1/2} N^{3/2}}{12\sqrt{2}\pi} \xi\left(\mu\right).
\ee

This conjectural large $N$ dual of the $\CN=3$ theory makes two quantitative predictions for the strongly coupled $\CN=3$ Chern--Simons--matter theory with flavors. 
First of all, the free energy of the $\CN=3$ theory on $\IS^3$ is given by the same expression that the one for the ABJM theory \cite{ejm,dmp} but with the new Newton's 
constant (\ref{newnew}), 
\be
\label{n3fe}
-F_{\CN=3}(\IS^3)=\frac{\pi}{2 G_4}={\pi {\sqrt{2}} \over 3} N^{3/2} k^{1/2} \xi\left(\mu \right)
\ee
Another quantity that we are interested in is the VEV of supersymmetric Wilson loops. As usual in the AdS/CFT correspondence, this can be calculated by evaluating the 
regularized area of a fundamental string in the type IIA reduction of the above M-theory background. The resulting geometry, which includes the full backreaction of the D6 branes, is a warped compactification and we have not performed such a calculation. However, it was pointed out in \cite{gj} that
\be
R_{\rm str}^2 \sim \frac{1}{4}{R^3_{X_7}  \over N_f+k} = {2 \pi N^{1/2} \over {\sqrt{ N_f + 2k}}}
\ee
and we then expect
\be
\label{expwl}
\left \langle W_{\tableau{1}} \right\rangle \sim \exp\left( {2 \pi N^{1/2} \over {\sqrt{ N_f + 2k}}} \right)
\ee
for both the $1/2$ and $1/6$ BPS Wilson loops. This VEV incorporates the 
screening effect on Wilson loops due to unquenched flavor. Indeed, we see from (\ref{expwl}) that, as the number of flavors grows, the exponent decreases. 
This might be interpreted as a conformal avatar of the screening effect. Also notice that, when $N_f\rightarrow 0$, one recovers in (\ref{expwl}) the right value for the ABJM limit. Notice that, in the computation leading to (\ref{expwl}), a possible contribution to the vev of strings ending on the D6 branes has not been taken into account. We will see however in this paper that a gauge theory computation at strong coupling leads to a result in agreement with (\ref{expwl}), thus indicating that such contributions are absent or subleading. 

\sectiono{Strong coupling limit and tropical geometry}

The ABJM matrix model can be solved in the planar limit simply by borrowing the exact solution of 
the lens space matrix model worked out in \cite{akmv,hy}. Let us now review some ingredients of this solution 
following the notations of \cite{dmp}.

\FIGURE{
\includegraphics[height=4.5cm]{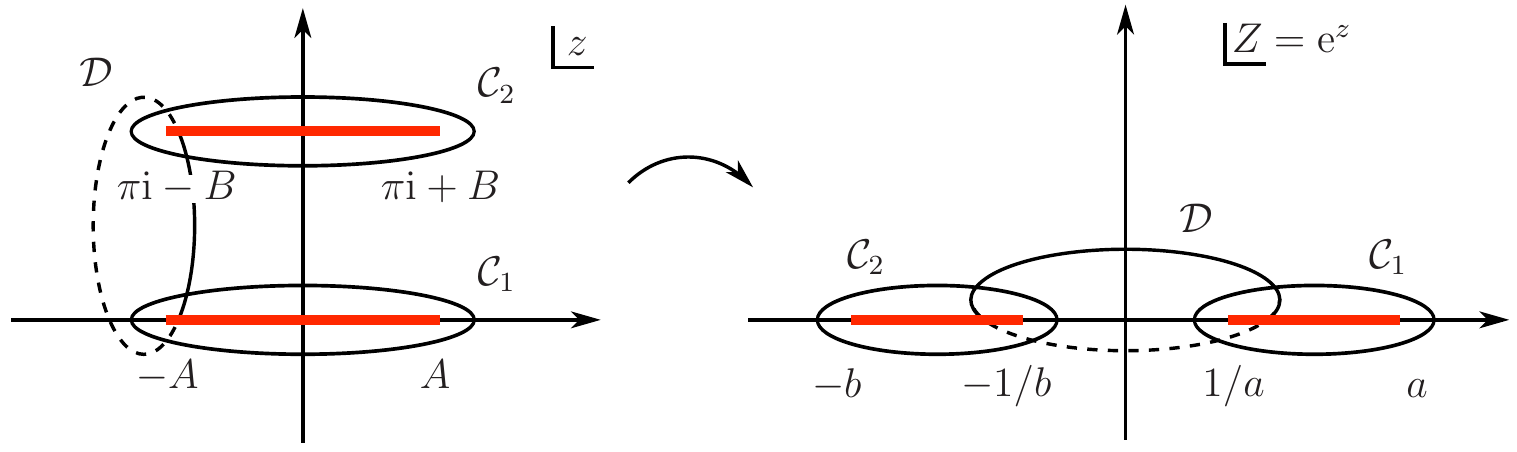} 
\caption{Cuts in the $z$-plane and in the $Z$-plane.}
\label{cuts}
}

At large $N$, the two sets of eigenvalues, $\mu_i$, $\nu_j$ in (\ref{intdef}) condense 
around two cuts. The cut of the $\mu_i$ eigenvalues is centered around $z=0$, while that of the 
$\nu_i$ eigenvalues is centered around $z=\pi \ri$. We will write these cuts as
\be
\label{ABcuts}
\CC_1=(-A, A), \qquad \CC_2=(\pi \ri -B, \pi \ri +B),
\ee
in terms of the endpoints $A, B$. It is also useful to use the exponentiated variable 
\be
Z=\re^z, 
\ee
In the $Z$ plane the cuts (\ref{ABcuts}) get mapped to 
\be
\label{Zend}
(1/a, a), \qquad (-1/b, -b) , 
\qquad\qquad
a=\re^A, \qquad b =\re^B, 
\ee
which are centered around $Z=1$, $Z=-1$, respectively, see \figref{cuts}. 
We will use the same notation $\CC_{1,2}$ for the cuts in the $Z$ plane. The total resolvent of the matrix model, 
$\omega(z)$ is defined as \cite{hy}
\be
\label{resolv}
\omega(z)
=g_s\left<\tr\left(\frac{Z+U}{Z-U}\right)\right>
=g_s \left\langle \sum_{i=1}^{N_1} \coth \left( {z-\mu_i \over 2} \right) \right\rangle +g_s \left\langle\sum_{j=1}^{N_2}  \tanh \left( {z-\nu_j \over 2} \right)\right\rangle
\ee
where $U$ is given in (\ref{U}). 

We will denote by $\omega_0(z)$ the planar limit of the resolvent, which was found in explicit form in \cite{hy}. It reads, 
\be
\label{explicitRes}
\omega_0(Z)  {\rd Z \over Z} = \log \biggl( {1 \over 2 \beta } \Bigl[ f(Z) -{\sqrt{ f^2(Z) -4 \beta^2 Z^2}}  \Bigr]\biggr){\rd Z \over Z} ,  
\ee
where
\be
f(Z)=Z^2-\zeta Z +1.
\ee
Here, $\beta$ and $\zeta$ are ``bare" parameters characterizing the resolvent. If we write
\be
\label{sigmaz}
\sigma(Z)= f^2(Z) -4 \beta^2 Z^2 =\left(Z-a\right) \left(Z-1/a\right) \left(Z+b\right) \left(Z+1/b\right) 
\ee
we see that the parameters $\zeta, \beta$ are related to the positions of the endpoints of the cuts as follows
\be
\label{zetadef}
\zeta={1\over 2}\left( a  +{1\over a} -b -{1\over b}\right), 
\qquad 
\beta={1 \over 4}\left( a  +{1\over a} +b +{1\over b}\right).
\ee

All the relevant planar quantities can be expressed in terms of period integrals of the one-form $\omega_0(z)\rd z$. The 't~Hooft parameters 
\be
t_i=g_s N_i, \qquad i=1, 2, 
\ee
are given by the period integrals
\be
\label{tperiods}
t_i ={1\over 4\pi \ri} \oint_{\CC_i} \omega_0 (z)  \rd z , \qquad i=1,2.
\ee
The total 't Hooft parameter 
\be
t=t_1+t_2
\ee
is related to the modulus $\beta$ as
\be
\label{ABJMbeta}
\beta=\re^t. 
\ee
The planar free energy $F_0$ satisfies the equation
\be
\label{pder}
\CI\equiv
{\partial F_0 \over \partial t_1} -{\partial F_0 \over \partial t_2}-\pi\ri t
=-\frac{1}{2}\oint_{{\cal D}} \omega_0(z) \rd z, 
\ee
where the ${\cal D}$ cycle encloses, in the $Z$ plane, the interval between $-1/b$ 
and $1/a$, see \figref{cuts}. Another quantity that can be computed as a period integral is the 1/6 BPS Wilson loop, since \cite{mp}
\be
\left \langle W^{1/6}_{\tableau{1}} \right\rangle_0
=\oint_{\CC_1}\frac{\rd Z}{4\pi \ri}\,\omega(Z).
\label{1/6-integral}
\ee
The 1/2 BPS Wilson loop is then given by 
\be
\left \langle W^{1/2}_{\tableau{1}} \right\rangle_0
=\oint_{\CC_1}\frac{\rd Z}{4\pi \ri}\,\omega(Z)+\oint_{\CC_2}\frac{\rd Z}{4\pi \ri}\,\omega(Z)=\oint_{\infty} {\rd Z \over 4 \pi \ri} \omega_0(Z).
\label{1/2-integral}
\ee
We finally recall that the solution to the original ABJM matrix model can be obtained by simply setting 
\be
\label{abjmdic}
t_1 =2 \pi \ri \lambda_1, \qquad t_2=-2\pi \ri \lambda_2, 
\ee
where
\be
\lambda_i={N_i \over k}
\ee
are the 't Hooft parameters of the ABJM model.

In order to make contact with the algebro-geometric formalism based on the spectral curve, notice that if we denote  
\be
\log \, Y=\omega_0(Z)
\ee
we find the following equation for an algebraic curve in $\IC^*\times \IC^*$:
\be
\label{mcurve}
Y+\frac{Z^2}{Y}-\sqrt{\frac{z_1}{z_2}}\left(Z^2-\frac{1}{\sqrt{z_1}}Z+1\right)=0, 
\ee
where the parameters $z_1, z_2$ are related to $\zeta, \beta$ as 
\be
\label{zxi}
\zeta={1\over {\sqrt{ z_1}}}, \qquad \beta= {\sqrt{z_2\over z_1}}.
\ee
The curve (\ref{mcurve}) is the mirror curve to the toric Calabi--Yau manifold known as local $\IP^1 \times \IP^1$, and the parameters $z_1, z_2$ become 
complex coordinates for the moduli space of the mirror Calabi--Yau.

In order to make contact with the AdS dual, one has to calculate the gauge theory/matrix model quantities at strong coupling. In \cite{mp, dmp} this was done essentially by 
computing exact interpolating functions at all couplings and then going to the strong coupling regime. However, the calculation of interpolating functions might become 
hard, specially in more complicated generalizations of ABJM theory like the theories with matter considered in \cite{hk,gj,taka}. In particular, one would like to 
have a computational framework to do calculations directly at strong coupling, without going through the determination of exact interpolating functions. 
We will now propose such a framework, and we will illustrate it by considering the ABJM theory. 

As explained in \cite{dmp}, the strong coupling 
limit of the ABJM theory corresponds to the large radius limit of this Calabi--Yau moduli space. This is the limit where 
\be
\label{lrlimit}
z_1, \,\, z_2 \rightarrow 0.
\ee
It is well-known that in this regime there are two natural period integrals $T_1, T_2$ given by
\be
\label{mirrormapone}
\ba
-T_1&= \log z_1 + \omega^{(1)} (z_1, z_2),\\
-T_2&= \log z_2 + \omega^{(1)} (z_1, z_2),
\ea
\ee
where 
\be
\omega^{(1)} (z_1, z_2) = 2\sum_{k,l\ge 0, \atop (k,l)\not=(0,0)} { \Gamma(2k + 2l) \over \Gamma(1+k)^2 \Gamma(1+l)^2} z_1^k z_2^l =2z_1 + 2z_2 + 3 z_1^2 + 12 z_1 z_2 + 3 z_2^2 + \cdots
\ee
We then see that in the limit (\ref{lrlimit}) one has
\be
\label{troplimit}
{\rm Re}\left(T_i\right) \rightarrow \infty, \qquad i=1, 2, 
\ee
and we can write, up to exponentially suppressed corrections, 
\begin{equation}
 z_1\approx \re^{-T_1}, \quad z_2\approx \re^{-T_2}\,.
\end{equation} 
The 't Hooft parameters of the ABJM model are large in this limit, and they behave like \cite{mp,dmp}
\be
\label{largelimit}
\lambda_i \sim {T_1 T_2 \over 8 \pi^2}, \qquad i=1,2.
\ee

We now ask the following question: what is the behavior of the planar resolvent, or equivalently the spectral curve (\ref{mcurve}), in this limit? Notice that 
the coefficients of (\ref{mcurve}), regarded as an equation for an algebraic curve, become exponentially large or small (or they remain constant). This kind of behavior has been 
very much studied recently in the mathematical literature and it is known as the {\it tropical limit} of the algebraic curve (or the {\it ultradiscretization} of the algebraic curve), 
see for example \cite{mikh-trop, iwao-integr}. 
This limit is only non-trivial if we scale $z,y$ in the same way, where 
\be
z=\log\, Z, \qquad y=\log \, Y, 
\ee
i.e. we have to consider the limit in which 
\be
{\rm Re}(z), \, \, {\rm Re}(y) \rightarrow \infty
\ee
as well.  For generic values of $z,y$ in this regime there is only one dominating term in (\ref{mcurve}), and the equation cannot be satisfied. To have a nontrivial equation we need at least two dominating terms which cancel each other. This gives us a set of linear equations on ${\rm Re}(z)$ and ${\rm Re}(y)$. Therefore, the ``ultradiscrete" limit of the curve can be represented as a collection of segments in the real plane. On each of these segments there is a linear relation between $z$ and $y$. It is an easy exercise to show that, for our particular example (\ref{mcurve}), the resulting diagram can be represented as in \figref{tropicalcurve}\footnote{In writing the linear equations for the segments, we have 
neglected constant imaginary parts, which are small in the tropical limit.}. This diagram is called a {\it tropical curve}. 

 \FIGURE[ht]{
\leavevmode
\centering
\includegraphics[height=7cm]{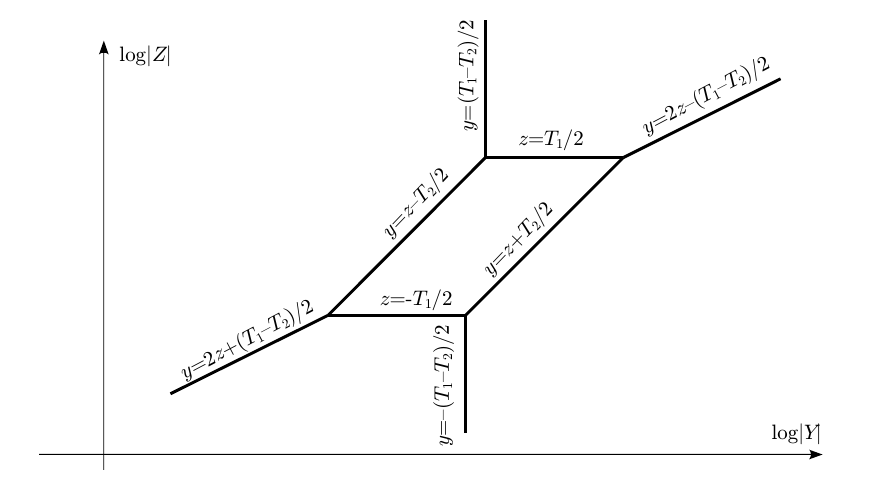}
\caption{The strong coupling limit of the curve (\ref{mcurve}) can be represented as a set of segments where the relation between $z$ and $y$ is linear. 
This limit is nothing but the ``ultradiscretization" or ``tropicalization" of the spectral curve.}
\label{tropicalcurve}
}
This two-dimensional plane can be understood as the base of the fibration 
\be
\ba
\mathbb{C}^*\times\mathbb{C}^*& \rightarrow\mathbb{R}^2\\
(Z,Y)&\mapsto (\log|Z|,\log|Y|).
\ea
\ee
The fiber is $\mathbb{S}^1\times \mathbb{S}^1$ and it is parametrized by the imaginary parts of $(z,y)$. A linear relation of the form 
$mz=ny+c,\;m,n\in\mathbb{Z}$ gives a line in $\mathbb{R}^2$ with a fiber $\mathbb{S}^1\subset\mathbb{S}^1\times\mathbb{S}^1$ with winding number $(n,m)$. Thus the lines in the picture correspond, in the original curve, to thin tubes connected at the vertices. This type of picture is familiar from local mirror symmetry: as emphasized in \cite{akv}, the mirror curve of a toric manifold, like (\ref{mcurve}), can be regarded as the thickening of the toric diagram in which lines become cylinders or tubes. In the 
strong coupling or large radius limit, the tubes become thinner and we get back the toric skeleton, which is now interpreted as a tropical curve.

 \FIGURE[ht]{
\leavevmode
\centering
\includegraphics[height=3.5cm]{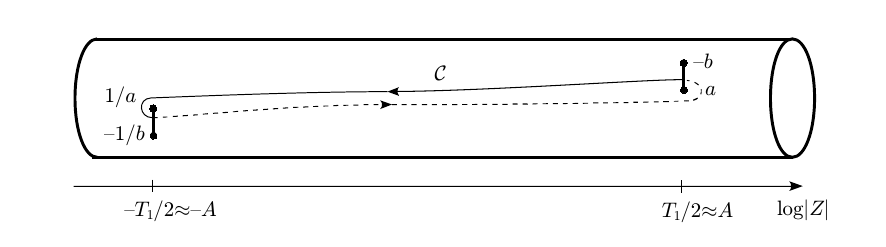}
\caption{The tropical limit of the cuts of the $Z$-plane, represented here as a cylinder.}
\label{tropcuts}
}

In order to do calculations at strong coupling we have to understand what happens to the period integrals in the regime (\ref{troplimit}). It can be shown rigorously 
(see for example \cite{iwao-integr}) that, in the limit (\ref{troplimit}), the periods of differentials on the original curve can be computed directly on the tropical curve, and they reduce to simple contour integrals along the two-dimensional diagram in \figref{tropicalcurve}. We then have to determine what is the tropical limit of the contours. 
To do this, we first note that in the limit (\ref{troplimit}) the endpoints of the cuts behave like
\be
\label{endbeh}
\ba
A& \approx \log\zeta \approx T_1/2\approx T_2/2,\\
a &\approx -b \approx\zeta\approx \re^{A},\\
{1\over a}& \approx  -{1\over b} \approx \re^{-A}.
\ea
\ee
Let us now represent the $\mathbb{C}^*$ domain of the variable $Z$ as an infinite cylinder, as in \figref{tropcuts}. This picture also shows 
the contour $\CC \equiv \CC_1$ around the cut $[1/a,a]$ in the tropical limit. Since our curve (\ref{mcurve}) 
is a double covering of this cylinder, we can build it 
out of two copies of $\mathbb{C}^*$ glued along the cuts shown in \figref{tropcuts}. To see this in detail, 
let us look at the diagram shown in \figref{tropicalcurve} and let us thicken it in order 
to reconstruct the original curve (\ref{mcurve}). If we remove the two horizontal segments, the thickening gives two infinite tubes which can be parametrized by $z$. Each of these tubes can be identified in turn with a copy of $\mathbb{C}^*$. In order to recover the full curve, we have to add the thickened horizontal segments. They give two horizontal tubes connecting the two copies of $\mathbb{C}^*$ at $z\approx-A$ and $z\approx A$. Notice that these locations are the positions of the small cuts drawn in \figref{tropcuts}. Since the solid and dashed pieces of $\CC$ depicted in \figref{tropcuts} lie on different copies of $\mathbb{C}^*$, we conclude that the contour $\CC$ around the cut $[1/a,a]$ becomes, in the tropical limit, the two-dimensional contour around the parallelogram shown in \figref{paral}.

\FIGURE[ht]{
\leavevmode
\centering
\includegraphics[height=7cm]{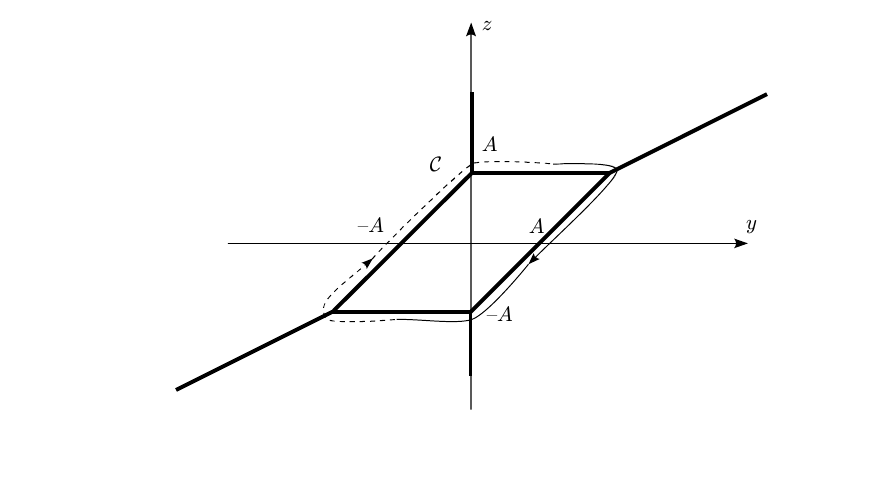}
\caption{The contour $\CC$ around $[1/a,a]$ becomes a parallelogram around the tropical curve.}
\label{paral}
}

Let us now consider the contour ${\cal D}$ in \figref{cuts}, which encircles the cut $[-1/b,1/a]$. This cut corresponds to the horizontal tube at $z\approx -A$, therefore the 
contour becomes a non-trivial cycle around the tube. In the tropical limit it can be schematically drawn as in \figref{dcontour}. 

\FIGURE[ht]{
\leavevmode
\centering
\includegraphics[height=7cm]{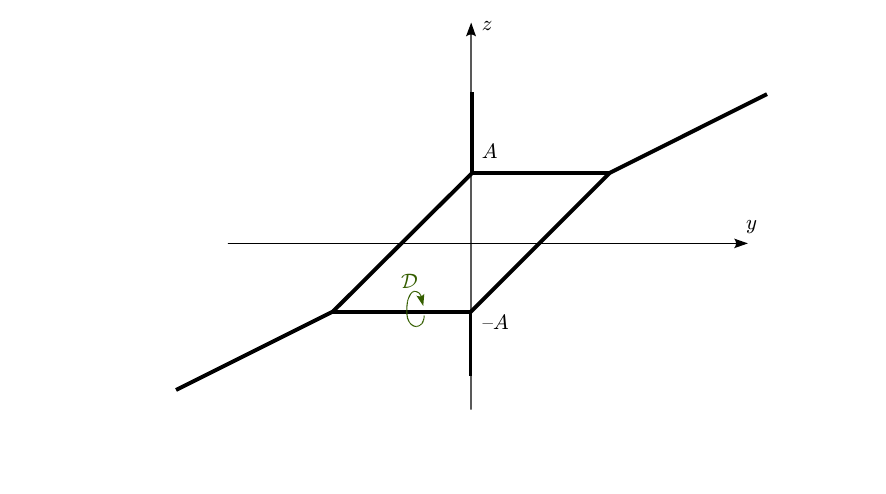}
\caption{The contour ${\cal D}$ around $[-1/b,1/a]$ becomes a non-trivial cycle around the shrinking tube.}
\label{dcontour}
}

We can now use this formalism to compute some interesting physical quantities at strong coupling. For simplicity we will restrict ourselves to the ABJM slice 
$N_1=N_2$. The resolvent is 
\be
\omega_0(z) =y(z)\,  \rd z. 
\ee
We first determine the relation between the 't Hooft parameter and the modulus $A$ as follows:
\be
 2\pi \ri\lambda=t_1=\frac{1}{4\pi \ri}\oint_\mathcal{C} y(z) \, \rd z \approx -\frac{A^2}{\pi \ri},
 \ee
which leads to 
\be
\label{Aprox}
A \approx \pi {\sqrt{2 \lambda}}
\ee
in agreement with the result (\ref{largelimit}) from \cite{mp,dmp}. The 1/6 BPS Wilson loop is given by a integral around the contour $\CC$:
\be
\label{wlf}
\left \langle W^{1/6}_{\tableau{1}} \right\rangle_0=\oint_{\CC} {\rd  z \over 4 \pi \ri } \re^z y(z). 
\ee
In the tropical limit, this becomes an elementary integral around the parallelogram shown in \figref{paral}. We then find, 
\be
\left \langle W^{1/6}_{\tableau{1}} \right\rangle_0 \approx  
 {1\over 4 \pi \ri} \left( \int_{-A}^{A} \left(z-A\right) \re^z\,  \rd z -\int_{-A}^{A } \left(z+ A\right) \re^z\,  \rd z \right) 
 \approx {\ri A\over 2  \pi } \re^A.
\ee
Using (\ref{Aprox}) we find
\be
\left \langle W^{1/6}_{\tableau{1}} \right\rangle_0 \approx {\ri \over 2}  \sqrt{2 \lambda} \re^{\pi {\sqrt{2 \lambda}}}
\ee
which is the result obtained in \cite{mp} from the exact interpolating function, 
up to an overall phase (this is due to the fact that we neglected constant subleading imaginary pieces in the equations for the segments of the tropical 
curve). 

Another quantity that we can compute is the free energy at strong coupling. From (\ref{pder}) we find, 
\be
\label{monint}
\frac{\partial F_0}{\partial \lambda}=-{\pi \ri}\oint_\mathcal{D}y \, \rd z ={\pi \ri}\oint_\mathcal{D}z \, \rd y \approx -\pi \ri A\, {\rm mon}_{\mathcal{D}} \, y,
\ee
where $\mon_\mathcal{D}$ denotes the monodromy along the cycle $\mathcal{D}$. Since $a\sim b\rightarrow\infty$ and $Z\sim 1/a\sim 1/b$, we have
\begin{equation}
\label{monypure}
 y \approx 2\log\left\{\sqrt{Z-1/a}-\sqrt{Z+1/b}\right\}+\mathrm{const}., 
\end{equation} 
and 
\be
{\rm mon}_{\mathcal D }\, y=2\pi \ri.
\ee
We conclude that
\be
\frac{\partial F_0}{\partial \lambda}\approx 2 \pi^2 A \approx 2 \pi^3 {\sqrt{2\lambda}}
\ee
which is the result obtained in \cite{dmp}. Of course, the interest of this tropical formalism is the generalization to 
more complicated situations. This we will do in the next section, where we will consider the ABJM theory with fundamental matter introduced in section 2.

\sectiono{Quenched flavor in Chern--Simons--matter theories}

\subsection{The quenched approximation in the matrix model}

In studying theories with fundamental matter 
multiplets in the context of the AdS/CFT correspondence, there have been essentially two approaches. In the first one, called the {\it quenched} or the {\it probe} approximation, one assumes that the number of flavors is much smaller than the number of colors. Since the flavor multiplets are usually obtained by adding branes 
to the original theory, the quenched approximation is equivalent to treating these branes as probes, and one assumes that they do not backreact on the background (see \cite{quenchedrev} for a review and a list of references for this approach). One can go beyond the 
quenched approximation and consider {\it unquenched} flavor, where the full backreaction of the branes is taken into account, see \cite{ramallo} for a recent review with
references. It turns out that these two approaches have counterparts in the study of the matrix model (\ref{gjmm}) including flavor multiplets. We will first set up the matrix model analogue of the quenched approximation, and we will consider the full unquenched theory in the next 
section. 

In the matrix model (\ref{gjmm}), the inclusion of fundamental flavors leads to the insertion of two determinant-like operators 
\be
\label{opin}
\ba
 & \prod_{i=1}^{N_1} \left(  2 \cosh {\mu_i \over 2} \right)^{-N_f^{(1)}}  \prod_{j=1}^{N_2} \left(  2 \cosh {\nu_j \over 2} \right)^{-N_f^{(2)}}\\
 &  =\exp\left[ - N_f^{(1)} \sum_{i=1}^{N_1} \log \left(  2 \cosh {\mu_i \over 2} \right) - N_f^{(2)} \sum_{j=1}^{N_2} \log \left(  2 \cosh {\nu_j \over 2} \right) \right].\ea
 \ee
 We can treat these insertions as operators which perturb the partition function without changing the 
spectral curve or resolvent of the ABJM theory. To see how this works in practice, we write the partition function (\ref{gjmm}) as a normalized vev in the ABJM theory, 
\be
Z_{\CN=3}(N_1, N_2, N_f ^{(1)}, N_f^{(2)}, g_s)=\langle \re^{-\CW} \rangle_{\text{ABJM}} Z_\text{ABJM}(N_1, N_2, g_s)
\ee
where
\be
\CW= N_f^{(1)} \CW_1 + N_f^{(2)} \CW_2
\ee
and
\be
\label{wops}
\CW_1= \sum_{i=1}^{N_1}  \log \left[2  \cosh {\mu_i \over 2} \right], \qquad \CW_2= \sum_{j=1}^{N_2}  \log \left[2 \cosh {\nu_i \over 2} \right].
\ee
We can then calculate the free energy of the $\CN=3$ theory as a cumulant expansion, 
\be
\label{freequenched}
F_{\CN=3}(N_1, N_2, N_f ^{(1)}, N_f^{(2)}, g_s)=F_{\text{ABJM}}(N_1, N_2, g_s)+\sum_{k= 1}^{\infty} { (-1)^k\over k!}\langle \CW^k \rangle^{(c)}_{\rm ABJM}
\ee
where $(c)$ denotes as usual the connected vev. Since $\CW^k$ is a polynomial of degree $k$ 
in the number of flavours $N_f^{(i)}$, the above cumulant expansion is an expansion 
around $N_f^{(i)}=0$. Equivalently, we can introduce the Veneziano parameters \cite{veneziano}
\be
t_f^{(i)}=g_s N_f^{(i)}. 
\ee
The perturbative series (\ref{freequenched}) is an expansion in the Veneziano parameters around $t_f^{(i)}=0$, which is valid for 
\be
t_f^{(i)}\ll 1, 
\ee
or equivalently
\be
N^{(i)}_f\ll {\rm min}(N_1, N_2),
\ee
which corresponds indeed to a quenched approximation. Each 
term in this series is given by an integrated correlator in the ABJM theory, which is computed with the master field described by the resolvent (\ref{explicitRes}). 
Since the spectral curve is not changed, this is equivalent to neglecting the backreaction of the D-branes on the original geometry. Diagramatically, the genus $g$ correction 
to $\langle \CW^k \rangle^{(c)}_{\rm ABJM}$ gives the contribution of $k$ ``quark" loops to the genus $g$ free energy, where all gluon diagrams of genus $g$ have been resummed. 

A similar perturbative scheme can be constructed for the calculation of operators $\CO$ in the matrix model (like for example Wilson loops):
\be
\label{wilsonq}
\langle \CO \rangle_{\CN=3} ={ \langle \CO \re^{-\CW} \rangle_{\rm ABJM} \over \langle  \re^{-\CW} \rangle_{\rm ABJM}}=\langle \CO \rangle_{\rm ABJM} +\sum_{k=1}^{\infty} {(-1)^k \over k!} \langle \CO \CW^k \rangle^{(c)}_{\rm ABJM}
\ee
The operator vevs appearing in (\ref{freequenched}) and (\ref{wilsonq}) can be computed by using the connected correlation functions of the ABJM model. These are defined 
by 
\be
\label{conncorr}
W(Z_1, \cdots, Z_h)=\left\langle  \tr {1\over Z_1 - U} \cdots \tr {1\over Z_h - U}  \right\rangle^{(c)}
\ee
where $U$ is given in (\ref{U}). These correlators have a genus expansion 
\be
\label{gw}
W(Z_1, \cdots, Z_h)=\sum_{g=0}^{\infty} g_s^{2g-2+h} W_g (Z_1, \cdots, Z_h)
\ee
which can be computed systematically with the techniques started in \cite{ambjorn} and culminated in \cite{eynard,eo} (for the Chern--Simons matrix models analyzed in 
this paper, one has to consider the slightly modified version of these techniques considered in \cite{mmopen,bkmp}). Let us 
consider for example the operators 
\be
\CO_a(X) =\tr \, f_a(X), \quad \widehat \CO_b(Y)=\tr\, g_b(Y),  \qquad a=1, \cdots, h_1, \quad b=1, \cdots, h_2, 
\ee
where 
\be
X={\rm diag}\left(\re^{\mu_i} \right), \qquad Y={\rm diag}\left(\re^{\nu_i} \right). 
\ee
In this notation, the operators (\ref{wops}) are written as 
\be
\CW_1(X)=\tr \log\left( X^{1\over2}+ X^{-{1\over 2}} \right), \qquad \CW_2(Y)=\tr \log\left( Y^{1\over2}+ Y^{-{1\over 2}} \right).
\ee
We have then the following result for the connected correlators of these operators, 
\be
\label{conncorros}
\ba
& \langle \CO_1(X) \cdots \CO_{h_1}(X) \widehat \CO_1(Y) \cdots\widehat  \CO_{h_2}(Y) \rangle^{(c)}=\\ 
& \qquad \oint_{\CC_1} { \rd X_1 \over 2 \pi \ri} \cdots \oint_{\CC_1}  { \rd X_{h_1}  \over 2 \pi \ri} \oint_{\CC2}   {\rd Y_{1}  \over 2 \pi \ri} \cdots 
\oint_{\CC2}  {\rd Y_{h_2}  \over 2 \pi \ri} \, W (X_1, \cdots, X_{h_1}, Y_1, \cdots, Y_{h_2})  \\
& \qquad \qquad \cdot f_1(X_1 )\cdots f_{h_1} (X_{h_1})g_1(Y_1 )\cdots g_{h_2} (Y_{h_2}).
\ea
\ee
This leads to a systematic $1/N$ expansion by using (\ref{gw}). The planar limit of the one-point functions is given by the equivalent expressions
\be
\langle \CO(X) \rangle_0= t_1 \int_{1/a}^a \rho_1(\mu) f(\mu)  \rd \mu, \qquad 
\langle \CO(Y)  \rangle_0= t_2 \int_{-b}^{-1/b} \rho_2 (\nu) g(\nu) \rd \nu.
\ee

\subsection{Quenched expansion at weak coupling}

We will now present some concrete examples of the quenched approximation, calculated at weak coupling. The results can be tested with perturbative calculations in 
the matrix model. For simplicity, we will set $N_f^{(2)}=0$, and we will focus on the free energy.  The first order correction in $N_f^{(1)}$ to the planar free energy is given by 
\be
\label{firstcorr}
-\langle \CW_1\rangle_0=-\oint_{\CC_1} {\rd Z \over 2 \pi \ri}  \,  \log\left( Z^{1\over2}+ Z^{-{1\over 2}} \right) W_0(Z), 
\ee
where 
\be
W_0(Z)={1\over 2 Z}  \omega_0(Z).  
\ee
The second order planar correction is
\be
\label{secondcorr}
{1\over 2!} \langle \CW_1^2 \rangle^{(c)}_0={1\over 2!} \oint_{\CC_1} {\rd X_1\over 2 \pi \ri}  \oint_{\CC_1} {\rd X_2 \over 2 \pi \ri}   \log\left( X_1^{1\over2}+ X_1^{-{1\over 2}} \right) 
\log\left( X_2^{1\over2}+ X_2^{-{1\over 2}} \right) W_0(X_1, X_2)
\ee
and $W_0(X_1, X_2)$ is the two-cut, two-point planar correlator of the matrix model. It is related to the Bergmann kernel of the spectral curve $B(X_1, X_2)$ 
by \cite{eynard}
\be
\label{berrel}
W_0(X_1, X_2)= B(X_1, X_2) -{1\over \left(X_1 -X_2\right)^2} 
\ee
and it was first calculated by Akemann \cite{akemann} in the useful form:
\be
\ba
W_0(X_1,X_2)&={1\over 4(X_1 - X_2)^2} \biggl(   {\sqrt  {(X_1 - a) (X_1 - 1/a)(X_2 +b) (X_2 +1/b) \over (X_1 +b) (X_1 +1/b)(X_2 -a) (X_2 - 
                      1/a)}} \\
                      & \, \, \ \ \ \ \ \ \ \ \ \ \ \ +   {\sqrt  {(X_1 +b) (X_1 +1/b)(X_2-a) (X_2 - 1/a) \over (X_1 - a) (X_1 - 1/a)(X_2 +b) (X_2 + 
                 b)}} \biggr) \\ & 
                      + {(a+1/b)(b+1/a) \over 4 {\sqrt{\sigma(X_1) \sigma(X_2)}}}  {E(k) \over K(k)}  - {1\over 2
                      (X_1 - X_2)^2},
                      \ea
\ee
where $\sigma(Z)$ is given in (\ref{sigmaz}). 

An efficient way to calculate the above integrals at weak coupling is to perform the change of variables
\be
X={a-a^{-1} \over 2} y + {a+ a^{-1} \over 2}, 
\ee
and expand the integrand in series in $t_i$  around $t_i=0$. The coefficients of the resulting series are relatively simple integrals, which can be computed by 
deforming the contour in terms of residues at infinity. The result one obtains is 
\be
\label{exq}
\ba
\langle \CW_1 \rangle_0& ={t_1^2 \over 8} + {1\over 96}t_1^2 \left(t_1 + 6 t_2 \right) + {1\over 64} t_1^2 t_2^2 + {1\over 3072} t_1^2 t_2^2 \left( t_1^2 -12 t_1 t_2  + t_2^2\right) +\cdots,\\ 
{1\over 2!} \langle \CW_1^2 \rangle_0^{(c)}&={t_1^2 \over 64} + {1\over 64}t_1^2 t_2 -{1\over 6144}  t_1^2 \left( t_1^2 + 24 t_1 t_2 -48 t_2^2\right)+\cdots
\ea
\ee
This can be explicitly checked against a direct calculation of the matrix integral.  Indeed, we find in matrix model perturbation theory
\be
\ba
 F_{\CN=3}& =F_{\rm ABJM} - N_f^{(1)} \left[ g_s {N_1^2 \over 8}  +g_s^2 \left( {N_1^3 \over 96} -{N_1^2 N_2 \over 16} -{5 N_1\over 192}\right)  + g_s^3 \left( {N_1^2 N_2^2  \over 64} + {N_1^2 \over 192} \right) + \cdots\right] \\
&  +(N_f^{(1)}) ^2 \biggl[ g_s^2  {N_1^2 \over 64} - g_s^3 \left( {N_1^2 N_2  \over 64} + {N_1 \over 128}\right) \\ 
&  \qquad -g_s^4 \left( {N_1^4 \over 6144} -{N_1^3 N_2 \over 256} - {N_1^2 N_2^2 \over 128} -{11 N_1^2 \over 248}-{3 N_1 N_2 \over 512} \right) + \cdots \biggr] 
\ea
\ee
whose planar part agrees with (\ref{exq}). 

\subsection{Quenched expansion at strong coupling}

Since the correlation functions (\ref{conncorr}) are given by contour integrals of meromorphic differentials, we can compute them with the 
tropical techniques that we introduced in the last section. We will focus on the free energy on $\IS^3$, and we will assume for simplicity that 
$N_f^{(2)}=0$ so that $N_f^{(1)}=N_f$. We can write
\be
\label{contourfirstf}
-\langle \CW_1\rangle_0 =-{1\over 4\pi \ri g_s}\oint\limits_{\mathcal{C}}\omega_0(z) f(z) \rd z 
\ee
where
\be
f(z)=\log\left(2 \cosh\frac{z}{2} \right).
\ee
In the tropical limit in which $z$ is large the function $f$ simplifies as
\begin{equation}
\label{troppot}
 f(z) \approx \frac{|z|}{2}.
\end{equation} 
Then, the contour integral (\ref{contourfirstf}) becomes
\be
- \oint_{\mathcal{C}}\omega_0(z) f(z) \rd z \approx
 \int\limits_{-A}^{A}\frac{|z|}{2}\left(z+A\right)\rd z-\int\limits_{-A}^{A}\frac{|z|}{2}\left(z-A\right)\rd z=A^3 \approx  \left(2\pi^2\lambda\right)^{3/2} 
 \ee
and the first correction of order $\CO(N_f)$ to the free energy is 
\be
\label{sfirst}
-\frac{\pi}{4} N_f N\sqrt{2\lambda}\,.
\end{equation} 

The next order $\CO(N_f^2)$ is much harder to compute with this technique, but it still can be done. Since this correction involves the two-point correlation function, which is 
essentially equal to the Bergmann kernel of the curve, what we have to do is to find the tropical limit of this kernel. To do this, we first discuss the tropical limit of holomorphic forms.

In tropical geometry a tropical holomorphic 1-form is a locally constant real 1-form with a ``conservation'' condition in the vertices, and which is zero on the external legs (for basic notions of tropical geometry see e.g. \cite{mikh-trop}). The dimension of the space of holomorphic 1-forms is obviously equal to the number of independent cycles of the graph, which coincides with the genus of the complex curve. In our case this space is a one-dimensional space with a basis $h$ such that 
\be
h=\pm \rd z
\ee
on the left and right sides of the parallelogram in Fig. \ref{tropicalcurve}, respectively, and 
\be
h=\pm \rd y
\ee
on the upper and lower sides. We also have $h=0$ on the external legs. 

One can realize a tropical holomorphic 1-form as a limit of a complex holomorphic 1-form: as we have seen, for each edge of the tropical curve we have an integer direction vector $(n,m)$. Then one can associate the integral of a complex holomorphic 1-form around the corresponding tube with the value of the tropical one form on this vector. In this way, the ``conservation'' condition in the vertices is a trivial consequence of holomorphicity. Since the external legs of the graph correspond to marked points on the complex curve, the absence of poles of the complex holomorphic 1-form at these points corresponds to the condition that the tropical holomorphic 1-form is zero on the external legs. In our particular case one can show explicitly that, in the tropical limit,
\begin{equation}
\label{trophol}
 \frac{\rd Z}{\sqrt{\sigma(Z)}}\approx \re^{-A}h\,.
\end{equation} 
The general notions of tropical Jacobian, Abel-Jacobi map and theta function were introduced in \cite{mikh-theta}. In our case the tropical Jacobian of our tropical curve $C_\trop$ is just 
\be
J_\trop=\mathbb{R}/L\mathbb{Z}\cong \mathbb{S}^1
\ee
where 
\be
L=\oint_\mathcal{C}h=8A
\ee
is the perimeter\footnote{As usual in tropical geometry, the length of the edge is a ``geometric'' length with an extra weight $(n^2+m^2)^{-1/2}$.} of the parallelogram. The tropical version of the Abel-Jacobi map is 
\be
\ba
u_\trop:C_\trop &\rightarrow J_\trop\\
p &\mapsto u_\trop(p)=\int_{p_0}^ph \mod L\mathbb{Z}
\ea
\ee
 which equals the length of a path between the points $p$ and $p_0$. It can be obtained as a tropical limit of the ordinary Abel-Jacobi map:
 \be
 u(p)\approx \frac{\ri}{2\pi}u_\trop(p).
 \ee
 The tropical theta function (with an odd characteristic) is
\begin{equation}
 \theta_\trop(v)=\max_{n\in\mathbb{Z}}\left\{-\frac{1}{2}Ln^2+n(v-L/2)\right\}\;.
\end{equation}
One can easily show that
\begin{equation}
 \theta'_\trop(v)=\left[\frac{v}{L}\right]
\end{equation}
where $[\cdot]$ denotes the floor function. Thus
\begin{equation}
 \theta''_\trop(v)=\delta(v\,\, {\rm mod} L\mathbb{Z})\,.
\end{equation} 
This tropical theta function can be obtained as a limit of the ordinary theta function with an odd characteristic
\begin{equation}
 \Theta(z)=\sum_{n\in \mathbb{Z}} \re^{\pi \ri \tau n^2+2\pi \ri (z+\tau/2)}\,.
\end{equation} 
In the tropical limit we have 
\be
\tau\approx \frac{\ri L}{2\pi}\rightarrow \ri\infty,
\ee
 so one exponential will dominate the others in the sum. Thus one can deduce that
\begin{equation}
 \log\Theta\left(\frac{v}{2\pi \ri}\right)\approx \theta_\trop(v).
\end{equation} 
Any Bergmann kernel can be written as 
\begin{equation}
 B=B_\mathrm{sing}+B_\mathrm{hol}\,,
\end{equation} 
where $B_\mathrm{sing}$ is given by (see e.g. \cite{eo-algmeth})
\begin{equation}
 B_\mathrm{sing}(p_1,p_2)=\rd_1\rd_2\log\Theta(u(p_1)-u(p_2))
\end{equation} 
and $B_\mathrm{hol}$ is a holomorphic part. In our case it should be chosen such that
\begin{equation}
 \oint\limits_{\CC}B=0\,.\label{bergmann-cond}
\end{equation} 
In the tropical limit
\begin{equation}
 B_\mathrm{hol}(p_1,p_2)\approx \mathrm{const}\cdot h(p_1)h(p_2)\,,
\end{equation} 
while
\begin{equation}
\ba
 B_\mathrm{sing}(p_1,p_2)& \approx \rd_1\rd_2\theta_\trop(u_\trop(p_2)-u_\trop(p_1))=-\delta(u_\trop(p_2)-u_\trop(p_1))h(p_1)h(p_2) \\ 
 & \equiv
-h_\mathrm{diag}(p_1,p_2)\,,
\ea
\end{equation} 
where $h_{\mathrm{diag}}$ is supported on the diagonal and has the property
\begin{equation}
 \oint\limits_{\mathcal{C}\times\mathcal{C}}h_{\mathrm{diag}}(p_1,p_2)f(p_1,p_2)=\oint\limits_{\mathcal{C}}h(p)f(p,p).
\end{equation} 
Imposing the condition (\ref{bergmann-cond}) we get
\begin{equation}
 B(p_1,p_2)\approx B_\trop(p_1,p_2)=-h_\mathrm{diag}(p_1,p_2)+\frac{h(p_1)h(p_2)}{L}\,.
\end{equation} 

We can now compute the second order correction at order $\CO(N_f^2)$, (\ref{secondcorr}), by using tropical techniques. It is given by 
\be
\frac{N_f^2}{2!}\frac{1}{(2\pi \ri )^2}\oint\limits_{\mathcal{C}\times\mathcal{C}}B(p_1,p_2)\,\log\left( 2 \cosh\frac{z_1}{2}\right) \,\log \left(2 \cosh\frac{z_2}{2} \right), 
\end{equation}
since the double pole subtracted in (\ref{berrel}) does not contribute to the double contour integral. 
In the tropical limit this integral reads
\begin{equation}
\ba
  \oint\limits_{\mathcal{C}\times\mathcal{C}} B_\trop\left|\frac{z_1}{2}\right|\left|\frac{z_2}{2}\right|&=\frac{1}{2^5A}\left(\int_\mathcal{C}h|z|\right)^2-\frac{1}{2^2}\int\limits_\mathcal{C}h|z|^2=
\frac{1}{2^2}\left\{\frac{\left(6A^2\right)^2}{8A}-\frac{16A^3}{3}\right\}\\
&=-\frac{5}{3\cdot 2^3}\,A^3.
\ea
\end{equation} 
Using (\ref{Aprox}) we obtain that the correction of order $\CO(N_f^2)$ is
\begin{equation}
\label{ssecond}
{ 5\,N_f^2\pi {\sqrt{2}} \over 96}\lambda^{3/2}.
\end{equation} 
On the other hand, the AdS prediction for the free energy is given by (\ref{n3fe}). The quenched approximation is obtained by expanding 
this quantity for small $N_f$. Since
\be
\xi(\mu)=1-\sum_{k=1}^{\infty} {(1+ 2k) (2k-3)!! \over 4^k \, k! } (-\mu)^k=1+{\mu\over 4} -{5 \mu^2 \over 32} +\cdots
\ee
we find 
\be
F_{\CN=3}(\IS^3)=-{\pi {\sqrt{2}} \over 3} N^{3/2} k^{1/2} -{ \pi \over 4} N_f N \sqrt{2 \lambda} + {5\pi  {\sqrt{2}} \over 96} N_f^2 \lambda ^{3/2} +\CO(N_f^3). 
\ee
We then see that the tropical computations (\ref{sfirst}), (\ref{ssecond}) reproduce correctly the first two terms in this expansion. 

One can try to compute the next corrections by calculating the tropical limit of the connected correlators (\ref{conncorr}) for $h\ge 3$, but 
as we will see in the next section it is possible to solve the planar theory at all values of $N_f^{(i)}$ (i.e. in the Veneziano limit) 
and calculate the tropical limit directly. 

\sectiono{Unquenched flavor in Chern--Simons--matter theories}

We now solve the matrix model (\ref{gjmm}) in the planar limit, but for all values of $N_f^{(i)}$, by using the techniques of \cite{hy,suyama}. 

\subsection{Exact resolvent in the Veneziano limit}

The starting 
point in the calculation of the resolvent are the saddle-point equations
\be
\ba
{\mu_i \over g_s} + {N_f^{(1)} \over 2}   \tanh\left( {\mu_i \over 2} \right)=& \sum_{j\ne i}^{N_1}\coth\frac{\mu_i-\mu_j}2-\sum_{a=1}^{N_2}\tanh\frac{\mu_i-\nu_a}2, \\
-{\nu_a \over g_s}  + {N_f^{(2)} \over 2} \tanh \left( {\nu_i \over 2} \right)=& \sum_{b\ne a}^{N_2}\coth\frac{\nu_a-\nu_b}2-\sum_{i=1}^{N_1}\tanh\frac{\nu_a-\mu_i}2.
 \label{saddle}
\ea
\ee
We will solve instead the problem 
\be
\ba
\mu_i + { t_f^{(1)} \over 2} \tanh\left( {\mu_i \over 2} \right)=&{t_1 \over N_1} \sum_{j\ne i}^{N_1}\coth\frac{\mu_i-\mu_j}2+{t_2 \over N_2} \sum_{a=1}^{N_2}\tanh\frac{\mu_i-\nu_a}2, \\
\nu_a -{ t_f^{(2)} \over 2} \tanh\left( {\nu_i \over 2} \right)=&{t_2 \over N_2}  \sum_{b\ne a}^{N_2}\coth\frac{\nu_a-\nu_b}2+{t_1 \over N_1}\sum_{i=1}^{N_1}\tanh\frac{\nu_a-\mu_i}2,
 \label{saddlealt}
\ea
\ee
analytically in the parameters $t_{1,2}$ and $t_f^{(1,2)}$, and then we will perform the analytic continuation 
\be
\label{chsign}
t_2 \rightarrow -t_2. 
\ee
The procedure to solve this type of equations is as in \cite{akmv,hy,suyama}. We first introduce exponentiated variables
\be
Z_i = \re^{\mu_i}, \qquad W_a = \re^{\nu_a}. 
\end{equation}
In terms of these variables the saddle--point equations read
\be
\ba
\log Z_i + { t_f^{(1)} \over 2}  {Z_i-1 \over Z_i+1}
=& t_1\frac{N_1-1}{N_1}+t_2+\frac{2t_1}{N_1}\sum_{j\ne i}^{N_1}\frac{Z_j}{Z_i-Z_j}
    -\frac{2t_2}{N_2}\sum_{a=1}^{N_2}\frac{W_a}{Z_i+W_a}, \\
\log W_a -  { t_f^{(2)}  \over 2} {W_a-1 \over W_a+1}
=& t_1+t_2\frac{N_2-1}{N_2}+\frac{2t_2}{N_2}\sum_{b\ne a}^{N_2}\frac{W_b}{W_a-W_b}
    -\frac{2t_1}{N_1}\sum_{i=1}^{N_1}\frac{Z_i}{W_a+Z_i}. 
    \label{GT2-2}
\ea
\ee
 The resolvent $\omega_0(Z)$ is defined as in (\ref{resolv}), and it will have two cuts corresponding to the set of eigenvalues. Let $[a,b], [c,d] \subset \mathbb{R}$ be the cuts corresponding to $-W_a$ and $Z_i$, respectively. 
In terms of the planar resolvent we have, 
\be
\ba
\log Z + { t_f^{(1)} \over 2}  {Z-1 \over Z+1}=&{1\over 2} \left(  \omega_0(Z+\ri 0)+\omega_0(Z-\ri 0)\right), \\
\log (-W) -{ t_f^{(2)} \over 2} {W+1 \over W-1}=&{1\over 2} \left(  \omega_0(W+\ri 0)+\omega_0(W-\ri 0) \right).
\ea
\ee
As in \cite{suyama} we now define the functions, 
\begin{equation}
\ba
F(Z) &= \sqrt{\sigma(Z) }\int_c^d \rd X\,\frac{f(X)}{Z-X}, \\
G(Z) &= \sqrt{\sigma(Z)}\int_a^b \rd X\,\frac{g(X)}{Z-X}, 
\ea
\end{equation}
where
\begin{equation}
\ba
f(x) &= {1\over \sqrt{|\sigma(x))|}}\left(\log x + { t_f^{(1)}  \over 2} {x-1 \over x+1}\right),\\
g(x)&=-{1\over \sqrt{|\sigma(x))|}}\left(\log (-x) -{ t_f^{(2)}  \over 2} {x+1 \over x-1}\right).
\ea
\end{equation}
It is then easy to show that the planar resolvent, defined again by the VEV (\ref{resolv}), is given by
\be
\omega_0(Z)={1\over \pi}(F(Z) +G(Z)).
\ee
One can check that, as in the ABJM matrix model, 
\be
\label{epproduct}
ab=1, \quad cd=1.
\ee
This follows from the symmetry of the saddle point equations under $Z\rightarrow Z^{-1}$, $W \rightarrow W^{-1}$, together with 
the conditions on the endpoints imposed by the asymptotic behavior of $\omega_0(z)$ at infinity, 
\be
\label{asymres}
\omega_0(z) \sim t, \qquad z \rightarrow \infty.
\ee
From now on we will denote 
the two independent endpoints as $a,b$, as in the ABJM model, so that the cuts are $[1/a, a]$ and $[-b, -1/b]$. The two equations
\be
\label{thooftagain}
t_1=\oint_{C_1}\frac{\rd Z}{4\pi \ri}\frac{\omega_0(Z)}{Z}, \qquad t_2=\oint_{C_2}\frac{\rd Z}{4\pi \ri}\frac{\omega_0(Z)}{Z}
\ee
determine them $a,b$ as a function of $t_1, t_2$. Here, $\CC_i$ encircle $[1/a, a]$ and $[-b, -1/b]$, respectively.

We can now calculate the planar resolvent explicitly. It is given by two pieces. The first one is, 
\be
\int_{1/a}^a {\rd X \over Z-X}{  \log X \over \sqrt{|\sigma(X)|}} -\int_{-b}^{-1/b} {\rd X \over Z-X}{  \log \left(-X  \right) \over \sqrt{|\sigma(X)|}},
\ee
which is simply the resolvent of the lens space matrix model (\ref{explicitRes}). The second piece is
\be
 { t_f^{(1)} \over 2} \int_{1/a}^a \frac{\rd X}{Z-X}\frac{X-1}{X+1}\frac{1}{\sqrt{|\sigma(X)|}} 
+ { t_f^{(2)} \over 2} \int_{-b}^{-1/b} \frac{\rd X}{Z-X}\frac{X+1}{X-1}\frac{1}{\sqrt{|\sigma(X)|}}. 
\ee
These integrals can be expressed in terms of elliptic functions. In order to do so we use the results
\be
\ba
	\int_{1/a}^a \frac{\rd X}{(Z-X)\sqrt{|\sigma(X)|}}& = \frac{2 \sqrt{ab}}{1+ab}\frac{1}{(Z-a)(Z+b)}\left( (a+b)\Pi(n^+(Z),k)+(Z-a)K(k) \right), \\
	\int_{-b}^{-1/b}\frac{\rd X}{(Z-X)\sqrt{|\sigma(X)|}} &= \frac{2\sqrt{ab}}{1+ab}\frac{1}{(Z-a)(Z+b)}\left(-(a+b)\Pi(n^-(Z),k)+(Z+b)K(k)\right),
\ea
\ee
where
\be
	k^2=\frac{(a^2-1)(b^2-1)}{(1+ab)^2},Ê\quad n^+(Z)=-\frac{a^2-1}{1+ab}\frac{Z+b}{Z-a}, \quad n^-(Z)=-\frac{b^2-1}{1+ab}\frac{Z-a}{Z+b}.
\ee
Defining the auxiliary function
\be
\ba
&J(a,b,Z,s)=\\
& \frac{2\sqrt{ab}}{1+ab}\frac{1}{Z+s}\left(\left((a+b)\frac{Z-s}{(Z-a)(Z+b)}\Pi(n^+(Z),k)+\frac{Z-s}{Z+b}K(k)\right)-(Z\rightarrow -s)\right),
\ea
\ee
we finally obtain
\be
\label{matres} 
\omega_0(Z)=\omega_0^{\text{ABJM}} (Z)+\frac{\sqrt{\sigma(Z)}}{2 \pi}\left( t_f^{(1)} J(a,b,Z,1)+t_f^{(2)} J(-b,-a,Z,-1)\right),
\ee
where $\omega_0^{\text{ABJM}} (Z)$ is the resolvent in the theory without matter, and it is given in (\ref{explicitRes}). The asymptotic behavior (\ref{asymres}) 
determines
\be
\label{tVen}
\ba
& t=\log{(\beta)}\\
& -\frac{\sqrt{ab}}{\pi(1+ab)}\left(t_f^{(1)} \left((b+1)K(k)-(a+b)\Pi(n_a,k)\right)+t_f^{(2)} \left(-(a+1)K(k)+(a+b)\Pi(n_b,k)\right)\right),
\ea
\ee
where we have used the notation
\be
\label{ens}
n_a= \frac{1-a^2}{1+ab}, \quad n_b=\frac{1-b^2}{1+ab}.
\ee
The relation (\ref{tVen}) reduces to (\ref{ABJMbeta}) when both $t_f^{(1)}$ and $t_f^{(2)}$ go to zero.

\subsection{Weak coupling limit in the unquenched theory}

In order to test the above expressions, we can compute the expansion of the resolvent at weak 't~Hooft 
coupling (i.e. around $t_i=0$) but for arbitrary $t_f^{(i)}$. To do this, 
the first step is to express the endpoints of the cuts in terms of the 't Hooft parameters. The period integrals for $t_i$ can be expanded around 
$a=1$, $b=1$, and these series expansions can be inverted. At the first few orders in $t_i$ we find
\be
\ba
a &=1+  \frac{1}{\sqrt{T_f^{(1)}}} \: 2 \sqrt{t_1}+\frac{1}{T_f^{(1)}}\: 2 t_1+\frac{1}{6 \left(T_f^{(1)}\right)^{3/2}}\left(7-\frac{1}{T_f^{(1)}}\right) \: \frac{3}{2} t_1^{3/2}+\frac{1}{\left(T_f^{(1)}\right)^{3/2}}\: \frac{1}{2}\sqrt{t_1}t_2   \\
	&+ \frac{1}{2 \left( T_f^{(1)}\right)^2}\left(3- \frac{1}{T_f^{(1)}}\right) \: t_1^2+\frac{1}{\left(T_f^{(1)}\right)^2}\: t_1 t_2 +\cdots,  \\
b&=1+ \frac{1}{\sqrt{T_f^{(2)}}} \: 2 \sqrt{t_2}+\frac{1}{T_f^{(2)}}\: 2 t_2+\frac{1}{6 \left(T_f^{(2)}\right)^{3/2}}\left(7-\frac{1}{T_f^{(2)}}\right) \: \frac{3}{2} t_2^{3/2}+\frac{1}{\left(T_f^{(2)}\right)^{3/2}}\: \frac{1}{2}\sqrt{t_2}t_1   \\
	&+ \frac{1}{2 \left( T_f^{(2)}\right)^2}\left(3- \frac{1}{T_f^{(2)}}\right) \: t_2^2+\frac{1}{\left(T_f^{(2)}\right)^2}\: t_2 t_1+\cdots.
	\ea
	\ee
In these equations, 
\be
T_f^{(1)}= 1+{ t_f^{(1)}\over 4}, \qquad T_f^{(2)}=1-{t_f^{(2)}\over 4}. 
\ee
When $t_f^{(i)}=0$ we recover the mirror map at the orbifold point of \cite{akmv}. As a test of these results, we can calculate 
the coefficient of $1/Z$ in the resolvent. After the analytic continuation (\ref{chsign}), this coefficient computes the planar VEV of the supertrace of $U$ in the matrix model (\ref{gjmm}), as in (\ref{12wl}):
\be
\ba
& 2g_s \left  \langle {\rm Str}_{\tableau{1}} \, U \right \rangle_0= \zeta\\
	&+ \frac{t_f^{(1)}}{2\pi \sqrt{ab}(1+ab)}\left((1+ab)^2 E(k)+(b-a+a^2b+3ab^2)K(k)-4ab (a+b)\Pi(n_a,k)\right) \\
	&+ \frac{t_f^{(2)}}{2\pi \sqrt{ab}(1+ab)}\left((1+ab)^2 E(k)+(a-b+ab^2+3a^2b)K(k)-4ab (a+b)\Pi(n_b,k)\right),
	\ea
	\ee
	where (\ref{chsign}) must be implemented at the end of the calculation. 
At weak coupling we find, 
\be
\ba
\label{flavorstr}
& g_s \left\langle  {\rm Str}_{\tableau{1}}\, U \right \rangle_0= t_1 + t_2 + \frac{1}{2T_f^{(1)}} t_1^2 - \frac{1}{2T_f^{(2)}}t_2^2 \\
& +{1\over  4 \left( T_f^{(1)} \right)^2}\left( 1-{1\over 3 T_f^{(1)}}\right) t_1^3  -{t_1^2  t_2 \over 4  \left( T_f^{(1)} \right)^2} 
- {t_1t_2 ^2  \over 4  \left( T_f^{(2)} \right)^2} +{1\over  4 \left( T_f^{(2)} \right)^2}\left( 1-{1\over 3 T_f^{(2)}}\right) t_2^3+\cdots
\ea
\ee
The r.h.s. has the expected symmetries of the matrix model. Indeed, it is odd under
\be
t_1 \leftrightarrow -t_2, \qquad  t_f ^{(1)} \leftrightarrow -t_f^{(2)}.
\ee
 Notice that each term in the expansion (\ref{flavorstr}) is a rational function of the Veneziano parameters. 
 From the diagrammatic point of 
 view, each of these terms corresponds to a fixed planar ``gluon" diagram (with a boundary associated to the insertion of $U$) in which we have summed over all 
the ``quark" loops, i.e. the ``gluons" are quenched and 
the ``quarks" are dynamical.  One can actually check the first few terms written down in (\ref{flavorstr}) againts an explicit perturbative calculation in the matrix model. 

Based on \cite{dt}, one should expect that the VEV (\ref{flavorstr}) computes (twice) the VEV of the 1/2 BPS Wilson loop operator. However, in order to 
assert this one should first check that the construction of \cite{dt} of this operator extends to the flavored theory that we are considering here. In any case, the formulae (\ref{pder}), (\ref{1/6-integral}) remain valid in the flavored theory, 
since the 1/6 BPS Wilson loop operator can be constructed for all Chern--Simons--matter theories considered in \cite{kapustin}. We will now evaluate these formulae in the unquenched theory at strong coupling, by using tropical techniques. 

\subsection{Strong coupling limit in the unquenched theory}

For simplicity we will set $N_f^{(2)}=0$, $N_f=N_f^{(1)}$. We will write (\ref{matres}) as 
\begin{equation}
\label{matterres}
 \omega_0(z)=y(z)\,  \rd z, \quad y(z)=y_\pure (z)+\mu\, y_\matter(z).
\end{equation}
In this equation, $\mu$ is defined in (\ref{muxidef}), 
\begin{equation}
 y_\pure(z)=\omega^{\rm ABJM}_0(z)
\end{equation} 
is the equation of the spectral curve (\ref{explicitRes}) in the ABJM model, and
\begin{equation}
 y_\matter(z)=\frac{1}{2}\oint\limits_\mathcal{C}{\rd X \over 2 \pi \ri} \frac{1}{Z-X}\,\frac{X-1}{X+1}\,
 \frac{\sqrt{\sigma(Z)}}{\sqrt{\sigma(X)}} \label{y-matter-int}\,.
\end{equation} 
In the tropical limit $\zeta\approx \re^{A}$ is large, and we will set
\be
\beta=\re^K
\ee
where $K$ is a parameter to be determined. In the ABJM model with $N_1=N_2$ one has $K=0$. We will shortly determine the value of $K$ in the theory with unquenched flavor, in the 
tropical limit. We will assume that $A>0$, $|K|<A$, which will be justified {\it a posteriori}. 

The integral (\ref{y-matter-int}) can be evaluated ``tropically.'' To do this we use the tropical limit (\ref{trophol}) of the holomorphic 
one-form, as well as the limits
\begin{equation}
 \frac{Z}{Z-X}\approx \Phi(x,z)\equiv 
\begin{cases} 1,  &\text{if $x<z$}\\
   -\re^{-(x-z)}, &\text{if $x>z$}
 \end{cases}
 \ee
 and
 \be
 \frac{\re^{-A}\sqrt{\sigma(Z)}}{Z}\approx \pm\Xi(z)\equiv 
 \pm\begin{cases}  -\re^{-(x-z)},  &\text{if $z<-A$}\\
  1, &\text{if $-A<z<A$}\\
  -\re^{z-A}, &\text{if $z>A$}
 \end{cases}
\end{equation} 
\FIGURE[ht]{
\leavevmode
\centering
\includegraphics[height=7.5cm]{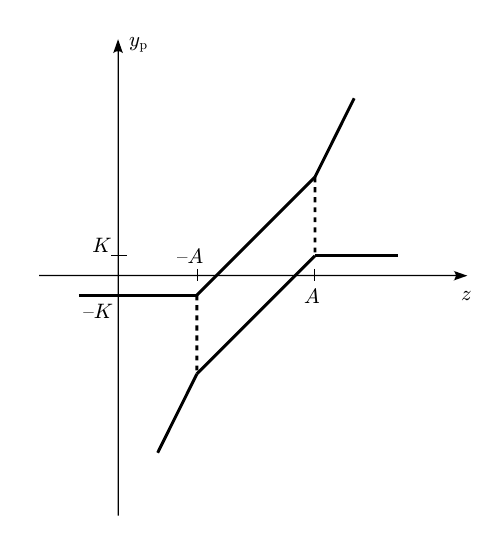} \qquad \includegraphics[height=7.5cm]{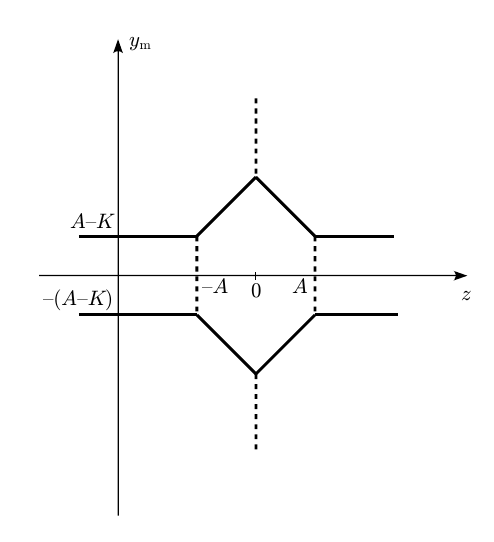}
\caption{The two-dimensional graphs representing the tropical limits of $y_\pure(z)$ (left) and $y_\matter(z)$ (right).}
\label{tropmatcurves}
}
Here the $\pm$ sign corresponds to the two determinations of the square root. One finally obtains
\begin{equation}
 y_\matter(z)\approx \pm\,\frac{1}{2}\,\Xi(z)\oint\limits_\mathcal{C}h_x \Phi(x,z) \sign x,
\end{equation} 
where $h_x$ is defined in (\ref{trophol}) but with $z$ replaced by $x$. 
In this result, $\sign x/2$ can be interpreted as the derivative of $|x|/2$, which is the tropical limit of the potential deformation, see (\ref{troppot}), 
and in principle we can generalized it to other deformations.
The tropical limit of $y_\matter(z)$ can be rewritten as
\begin{equation}
 y_\matter(z) \approx \pm\,\frac{1}{2} \int\limits_{\mathcal{C}_z}h_x \sign x=\pm  \begin{cases} A-K+(A-|z|),&\text{if $ -A<z<A$}, \\
  A-K, & \text{otherwise,} 
 \end{cases}
 \label{y-matter-trop}
\end{equation}
where $\mathcal{C}_z$ is a line connecting two different points on the curve with the same value of $z$. Equivalently one can obtain (\ref{y-matter-trop}) by taking the tropical limit in the explicit expression (\ref{matres}). The tropical limit of the two-valued functions $y_\pure(z)$, $y_\matter(z)$ can be represented by the two-dimensional graphs shown in \figref{tropmatcurves}. Of course, the diagram for $y_\pure(z)$ is 
nothing but the tropical curve represented in \figref{tropicalcurve}. 

We now want to find the relation between $A$ and $K$. To do this, we will impose for simplicity that $N_1=N_2=N$ in the $\CN=3$ Chern--Simons--matter theory. This means that the total 
't Hooft parameter $t=t_1 + t_2$ vanishes. It follows from (\ref{thooftagain}) that this sum can be evaluated by deforming the sum of the contours $\CC_1$ and $\CC_2$ to infinity and 
the origin, so we obtain
\begin{equation}
{\rm res}_{Z=\infty}\, \omega_0(Z) -{\rm res}_{Z=0}\, \omega_0(Z)=0,
\end{equation} 
which leads to
\be
K= {\mu \over 1+\mu} A.
\ee
Notice that in the limit $\mu=0$ we correctly reproduce $K=0$. With this relation at hand we can already add the two graphs to obtain 
the tropical curve representing the new resolvent $\omega_0(z)$, which is shown in \figref{tropflavor}. 
\FIGURE[ht]{
\leavevmode
\centering
\includegraphics[height=7.5cm]{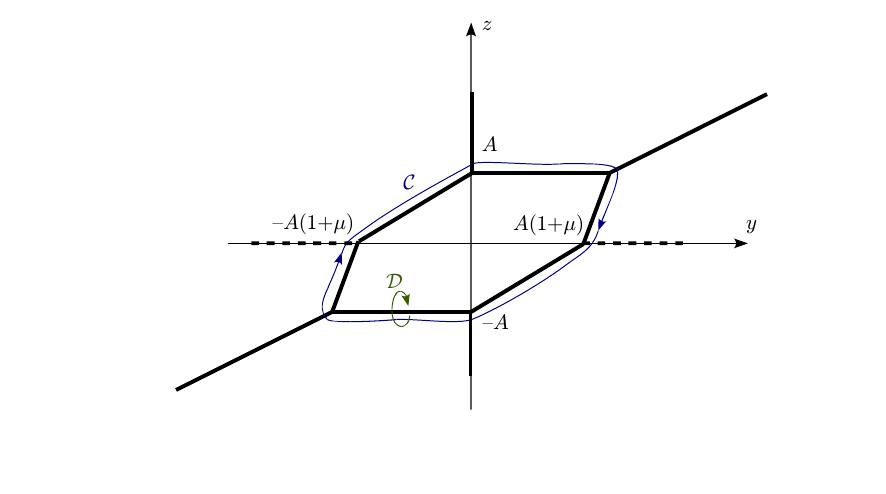}
\caption{The tropical curve representing the tropical limit of the resolvent (\ref{matres}) of the $\CN=3$ theory.}
\label{tropflavor}
}
The calculation of the different periods reduces, like before, to trivial line integrals on the plane. We first have to relate $A$ to the 't Hooft parameter. 
We have, for the period (\ref{thooftagain}), 
\begin{equation}
 2\pi \ri\lambda=t_1=\frac{1}{4\pi \ri}\oint_\mathcal{C} y(z) \, \rd z \approx -\frac{(1+\mu/2)A^2}{\pi \ri}
\end{equation}
and we find
\begin{equation}
\label{amatter}
A \approx { \pi {\sqrt{ 2 \lambda}} \over {\sqrt{1+\mu/2}}},
\ee
which is the deformation of the relationship (\ref{Aprox}) in the ABJM theory. For the planar free energy we have, as in the period integral 
(\ref{monint}) for the ABJM theory,
 \begin{equation}
 \frac{\d F_0}{\d\lambda}= -\pi \ri A\,{\rm mon}_\mathcal{D} y.
\end{equation} 
The monodromy of $y_\matter(z)$ can be computed as (\ref{monypure}). Indeed, 
the main contribution to the integral for $y_\matter$ is given by the vicinity of $1/a$, and we have
\be
\ba
  y_\matter & \approx \int\limits_{1/a}^\infty\frac{\rd X}{X-Z}\,\frac{\sqrt{(Z-1/a)(Z+1/b)}}{\sqrt{(X-1/a)(X+1/b)}}\\
& = \log\left\{\sqrt{Z-1/a}-\sqrt{Z+1/b}\right\}-\log\left\{\sqrt{Z-1/a}+\sqrt{Z+1/b}\right\}.
\ea
\ee
Then the monodromy around the cut $[-1/b,1/a]$ is
\begin{equation}
 {\rm mon}_{\mathcal{D}}y_\matter\approx 2\pi \ri\,.
\end{equation} 
We conclude that
\begin{equation}
\frac{\d F_0}{\d\lambda}\approx2\pi^2 A(1+\mu),
\end{equation}
or equivalently, 
\begin{equation}
F_0(\lambda) \approx \frac{1}{3}\pi^3\,\sqrt{2\lambda^3}\,\frac{1+\mu}{\sqrt{1+\mu/2}}.
\end{equation} 
This is in perfect agreement with the AdS prediction (\ref{n3fe}). Notice in particular that we have been able to reconstruct the 
full nontrivial function $\xi(\mu)$ involved in the volume of the tri-Sasakian target space (\ref{xitri}). 

Finally, we can calculate the vev of the $1/6$ BPS Wilson loop, which is given again by (\ref{wlf}) but now with the new resolvent. We obtain, 
\be
\ba
\left \langle W^{1/6}_{\tableau{1}} \right\rangle_0 &  \approx
 \frac{1}{4\pi i} \left(\int_{-A}^{A}\re^z\left(z-A-\mu(A-|z|)\right)\rd z -\int_{-A}^{A}\re^z\left(z+A+\mu(A-|z|)\right) \rd z \right)
 \\
& \approx
 \frac{\ri}{2}\,\sqrt{\frac{2\lambda}{1+\mu/2}} \exp\left( \pi\sqrt{\frac{2\lambda}{1+\mu/2}} \right).
\ea
\ee
With some more work one can show that the would-be $1/2$ BPS Wilson loop has the same leading, exponential dependence. This is 
in perfect agreement with the AdS prediction (\ref{expwl}), and the exponent should be equal to the regularized area of a fundamental string in the 
corresponding type IIA background. 

\sectiono{Conclusions}

The two main results of this paper are the following. First, we have shown that the strong coupling limit of Chern--Simons--matter theories corresponds, 
in the matrix models of \cite{kapustin}, to the tropical limit of the spectral curve. In this limit, the curve becomes a two-dimensional graph and the 
period integrals which give the main observables of the matrix model can be computed in a very simple way by evaluating line integrals along the graph. 
Second, we have obtained the planar resolvent of the matrix model describing the ABJM theories with flavor introduced in \cite{hk,gj,taka}, including all planar loops of 
fundamental matter. In other words, we have solved the Veneziano limit of these flavored theories. The unquenched regime at 
strong coupling can be studied by applying our tropical techniques to the exact resolvent, and we have reproduced in detail the AdS predictions of the 
conjectural large $N$ duals. In particular, we have reconstructed from a gauge theory calculation the volume of the tri-Sasaki Einstein manifolds involved 
in these AdS duals. 

There are many interesting problems which deserve further research. It would be interesting to analyze with our tropical techniques other Chern--Simons--matter theories, 
like for example the two-node quiver with arbitrary levels introduced and studied in \cite{gt}. The exact planar resolvent was recently written down in \cite{suyama}, and 
our techniques should lead to explicit expressions for its free energy at strong coupling. 
Since this theory has a conjectural large $N$ dual described by massive type IIA supergravity, one might be able to 
test directly in the matrix model many of the results recently discussed in \cite{ajtz}. 

On a more mathematical level, we have seen that the quenched approximation to the flavored theories studied in this paper involves 
in a natural way the tropical limit of the connected matrix model correlators (\ref{conncorr}). These correlators can 
be computed with the topological recursion of Eynard and Orantin \cite{eo}, and 
in section 4 we have described the tropical limit of one of the basic building blocks of this recursion, namely the tropical Bergmann kernel. 
It would be very interesting to see if the full topological recursion can be formulated directly in the tropical limit. This would 
give a powerful formalism to study perturbations of ABJM theory and it might find other applications in the future. 

It seems that, in the Chern--Simons--matter theories with less supersymmetry, the planar resolvent is much more complicated than in ABJM theory, and in particular 
we don't longer have the underlying local Calabi--Yau geometry that we had in (\ref{mcurve}). For example, the resolvent in (\ref{matres}) involves elliptic functions, in 
contrast to (\ref{explicitRes}). The tropical limit is however given by a graph which ressembles very much a blown-up toric Calabi--Yau manifold 
(see \figref{tropflavor}). It would be 
important to understand at which extent the solution to these Cherm--Simons--matter matrix models can be encoded in a toric Calabi--Yau.

\section*{Acknowledgments}
We would like to thank Davide Gaiotto, Daniel Jafferis, Igor Klebanov, Juan Maldacena, Dario Martelli, Carlos N\'u\~nez and Alfonso Ramallo 
for very useful discussions. 
The work of R.C.S. is supported by a grant from the Spanish Ministry of Science and Innovation, MICINN and FEDER (grant FPA2008-01838),
by Xunta de Galicia (Conseller\'\i a de Educaci\'on and grant PGIDIT10PXIB206075PR), and by the Spanish Consolider-Ingenio
2010 Programme CPAN (CSD2007-00042). The work of M.M. and P.P. is supported in part by the 
Fonds National Suisse. In addition, R.C.S. would like to thank the D\'epartement de Physique Th\'eorique of the University of Geneva for 
its hospitality. 


  \appendix
  \sectiono{Strongly coupled density of eigenvalues and tropical geometry}
  
  In this Appendix we rederive some of the results for the $\CN=3$ theory by using the approach of \cite{hklebanov}, and we compare it in detail to 
  our tropical methods. 
  
 The starting point of \cite{hklebanov} is an analysis of the ABJM matrix model (\ref{kapmm}) in the ABJM slice, at large $N$ but fixed $k$, which corresponds to the strongly coupled limit of the theory. This analysis makes possible to re-derive in a particularly simple and beautiful way the result of \cite{dmp}. Let us see how this is done, 
 following closely the steps in \cite{hklebanov}. The behaviour at large $N$ of the equilibrium eigenvalues of the matrix model is, as derived in \cite{dmp}, 
    \be
    \label{scaling}
  \mu_k =N^{1/2} x_k+ \ri \ell_k, \qquad \nu_k =N^{1/2} x_k-\ri \ell_k,  \qquad k=1, \cdots, N,
  \ee
  where $x_k$, $\ell_k$ are of order one at large $N$. At large $N$ the eigenvalues $x_k$, $\ell_k$ become dense, so that 
  \be
  {k\over N} \rightarrow \xi \in [0,1]
  \ee
  and they are described by the functions
\be
\rho(x) ={\rd \xi \over \rd x}, \qquad \ell(x). 
\ee
It is shown in \cite{hklebanov} that, when $N$ is large, the free energy of the matrix model can be written as 
\be
\label{hkfree}
-F=N^{3/2} \left[ {k \over \pi} \int \rd x \, x \rho(x) \ell(x) + \int \rd x \, \rho^2(x) f\left( 2 \ell(x)\right) -{m \over 2 \pi} \left( \int \rd x \, \rho(x)-1\right)\right]. 
\ee
Here, $f(t)$ is a periodic function of $t$, with period $2\pi$, and given by
\be
f(t) =\pi^2 -t^2, \qquad t\in [-\pi, \pi]. 
\ee
The last term in (\ref{hkfree}) involves, as usual, a Lagrange multiplier $m$ imposing the normalization of $\rho(x)$. 
 Notice that our sign convention for the free energy is the same as in \cite{dmp} and opposite to the one chosen in \cite{hklebanov}. Varying this functional w.r.t. $\rho(x)$ 
 and $\ell(x)$ one obtains the two equations
 \be
 \ba
 2 \pi \rho(x) f'\left(2 \ell(x)\right)&=-kx, \\
 4 \pi \rho(x) f\left(2 \ell(x)\right)&=m-2 k x \ell(x),
 \ea
 \ee
 which are solved by 
 \be
 \label{nomsol}
 \rho(x)={m \over 4 \pi^3}, \qquad \ell(x)={\pi^2 k x \over 2 m}. 
 \ee
The support of $\rho(x)$, $\ell(x)$ is the interval $[-x_*, x_*]$. One fixes $x_*$ and $m$ from the normalization of $\rho$ and by minimizing $-F$. This 
gives
\be
\label{xmABJM}
x_*=\pi {\sqrt{2 \over k}}, \qquad m={2 \pi^3 \over x_*}. 
\ee
Evaluating the free energy for the functions (\ref{nomsol}) and the values (\ref{xmABJM}) of $x_*$, $m$, one reproduces the result of \cite{dmp} for the free energy. 

The above results can be easily compared with our tropical analysis. The value of $x_*$ gives (up to a factor $N^{1/2}$) the position of the endpoint $A$, and it is in 
accord with the value of (\ref{Aprox}), since 
\be
\label{Axstar}
A =N^{1/2} x_*. 
\ee
The fact that the density $\rho(x)$ is constant follows from our result for the tropical limit of the curve. Indeed, the density of eigenvalues (normalized as to have an integral along the cut equal to one) is given by the well-known formula
\be
\rho(z) ={1\over 8 \pi^2 \lambda} {\rm disc}\, y(z)
\ee
where ${\rm disc}\, y(z)$ is the discontinuity of the curve through the cut $[-A, A]$. In our case 
this is just the constant $2A$, and it is given by the horizontal separation between the two 
diagonals in \figref{paral}. Changing variables from $z=N^{1/2}x$ to $x$ we find indeed, 
\be
\rho(x)= {\sqrt{2 k} \over 4 \pi}, 
\ee
in precise agreement with the result of \cite{hklebanov}. 

The inclusion of fundamental matter in the approach of \cite{hklebanov} is straightforward. $-F$ includes now the extra term 
\be
{ N^{3/2} N_f \over 2} \int \rd x \, \rho(x) |x|
\ee
which is the large $N$ limit of the operator in the exponential of (\ref{opin}) (as in (\ref{troppot})). 
The new saddle point equations are 
 \be
 \ba
 2 \pi \rho(x) f'\left(2 \ell(x)\right)&=-kx, \\
 4 \pi \rho(x) f\left(2 \ell(x)\right)&=m-2 k x \ell(x) -\pi N_f |x|,
 \ea
 \ee
with solution 
\be
\label{rhoma}
\rho(x) ={m - \pi N_f |x| \over 4 \pi^3}, \qquad \ell(x)= { k \pi^2 x \over 2\left( m-\pi N_f |x| \right)}.
\ee
Normalization of the density and minimization of $-F$ lead to
\be
\label{xmmatter}
x_*={2 \pi \over {\sqrt{ 2k + N_f}}}, \qquad m=2\pi^2 {k + N_f \over {\sqrt{ 2k + N_f}}}.
\ee
A straightforward calculation of $-F$ reproduces (\ref{n3fe}). 

Let us now compare this with the tropical approach. First of all, we have again the equality (\ref{Axstar}) between the endpoints of the cut in both approaches, involving now the value of $x_*$ obtained in (\ref{xmmatter}) and the value of $A$ obtained in (\ref{amatter}). The density of eigenvalues in the theory with matter can be obtained from the planar resolvent (\ref{matterres}) as 
\be
\rho(x)={N^{1/2} \over 8 \pi^2 \lambda} {\rm disc} \left( y_{\rm p}(z) + \mu y_{\rm m} (z)\right),
\ee
where $\mu$ is defined in (\ref{muxidef}). From \figref{tropmatcurves} and (\ref{y-matter-trop}) we read off immediately
\be
{1\over 2} {\rm disc} \, y_{\rm p}(z) =A-K, \qquad  {1\over 2} {\rm disc} \, y_{\rm m}(z) =2A -K -|z|, 
\ee
and we deduce
\be
\rho(x)={N^{1/2} \over 4 \pi^2 \lambda} \left[ (1+ \mu)A - \mu N^{1/2} |x| \right].
\ee
Plugging in the value of $A$ (\ref{amatter}), we recover precisely the form of $\rho(x)$ given in (\ref{rhoma}). This shows explicitly that 
the piece-wise linear densities obtained with the method in \cite{hklebanov} correspond to the tropical curves obtained in this paper. 

One advantage of the method of \cite{hklebanov} is that it gives the large $N$ limit of the free energy, at strong coupling, without 
directly using the resolvent of the model. Therefore, this method is useful when the resolvent is difficult to write down.
If on the contrary one is interested in calculating the resolvent of the model (to study for example weak coupling expansions), then the tropical approach 
developed in this paper provides a powerful method to extract the strong coupling limit from the resolvent. 
In addition, the method of \cite{hklebanov} assumes as an {\it ansatz} the scaling (\ref{scaling}) of the eigenvalues, while 
the approach based on studying the resolvent and its tropical limit provides a {\it bona fide} solution to the large $N$ theory without any 
further assumptions.

\end{document}